\documentclass[aps,pra,twocolumn,showpacs]{revtex4-1}
\usepackage{indentfirst}
\usepackage{bm}
\usepackage{verbatim}
\usepackage{graphicx}
\usepackage{subfigure}
\usepackage{epsfig}
\usepackage{amsmath}
\usepackage{booktabs} %�����߼Ӵ�
\usepackage{multirow}
\usepackage{txfonts} % Ĭ������times new roman
\usepackage{float}
\usepackage{tabularx}
\usepackage{threeparttable}
\usepackage[citecolor=blue,colorlinks=true,linkcolor=blue,urlcolor=blue,breaklinks=true]{hyperref}

\begin{document}
%\title{Protection of quantum coherence in a non-Hermitian driving quantum kicked rotor }%
%\title{Recovery of quantum coherence by non-Hermiticity in a noised quantum kicked rotor }%
\title{Adiabaticity in nonreciprocal Landau-Zener tunneling }%
\author{Wen-Yuan Wang$^{1,2}$}
\author{Bin Sun$^{3}$}
\author{Jie Liu$^{3,4}$}
\email[E-mail address: ]{jliu@gscaep.ac.cn.}
\affiliation{$^{1}$Key Laboratory of Atomic and Molecular Physics $\&$ Functional Materials of Gansu Province, College of Physics and Electronic Engineering, Northwest Normal University, Lanzhou 730070, China \\
$^{2}$Beijing Computational Science Research Center, Beijing 100193, China \\
$^{3}$Graduate School, China Academy of Engineering Physics, Beijing 100193, China \\
$^{4}$HEDPS, Center for Applied Physics and Technology, and College of Engineering, Peking University, Beijing 100871, China
}
\begin{abstract}
We investigate the Landau-Zener tunneling (LZT) of a self-interacting two-level system in which the coupling between the levels is nonreciprocal. In such a non-Hermitian system,
when the energy bias between two levels is adjusted very slowly, i.e., in the adiabatic limit, we find that a quantum state can still closely follow the eigenstate solution until it encounters the exceptional points (EPs) at which two eigenvalues and their corresponding eigenvectors coalesce. In the absence of the nonlinear self-interaction, we can obtain explicit expressions for the eigenvectors and eigenvalues and analytically derive the adiabatic LZT probability from invariants at EPs. In the presence of the nonlinear interaction, the dynamics of the adiabatic evolutions are explicitly demonstrated with the help of classical trajectories in the plane of the two canonical variables of the corresponding classical Josephson Hamiltonian.
We show that the adiabatic tunneling probabilities can be precisely predicted by the classical action at EPs in the weak nonreciprocal regime.
In a certain region of strong nonreciprocity, we find  that interestingly, the nonlinear interaction effects can be completely suppressed so that the adiabatic tunneling probabilities are identical to their linear counterparts. We also obtain a phase diagram for large ranges of nonreciprocity and nonlinear interaction parameters to explicitly demonstrate where the adiabaticity can break down,  i.e., the  emergence  of the nonzero tunneling probabilities even in adiabatic limit.
\end{abstract}
%\pacs{67.85.Hj, 05.45.Yv, 42.81.Dp} %PACS, the Physics and Astronomy
 \maketitle
\section{Introduction}
Landau-Zener tunneling (LZT) is a well-known phenomenon discussed in the quantum mechanics textbooks that  describes a system that goes from one side of an avoided level crossing to the other side at a certain sweeping rate \cite{Landau1932,Zener1932}. The LZT model has many important applications in various physical systems
demonstrated by recent progress in experiments, for example in superconducting qubits \cite{PhysRevLett.96.187002,PhysRevA.74.052330,Berns2008}, nitrogen-vacancy centers \cite{doi:10.1126/science.1181193}, quantum dots \cite{PhysRevLett.102.216802,doi:10.1126/science.1183628}, waveguide arrays \cite{PhysRevLett.94.113904}, and Bose-Einstein condensates \cite{PhysRevA.61.023402,PhysRevA.61.033603,PhysRevA.66.023404,PhysRevA.66.063603,PhysRevLett.87.140402,PhysRevA.65.063612,
PhysRevLett.90.170404,PhysRevLett.91.230406,PhysRevA.73.063609,PhysRevLett.96.020405,PhysRevLett.102.230401,Zenesini_2008,
PhysRevLett.103.090403,Chen2011,PhysRevLett.106.155302,Zenesini_2008,PhysRevLett.103.090403,Chen2011,
PhysRevLett.120.040407,PhysRevA.99.023616,PhysRevLett.125.213401}, to name only a few.

The LZT model has many extensions by taking diverse physical conditions into account, such as in multilevel systems \cite{PhysRevB.66.205303,PhysRevA.70.052708,Volkov_2004,PhysRevA.74.033414,PhysRevA.87.032701,PhysRevA.90.062509,
PhysRevA.94.042109, PhysRevA.96.022107,PhysRevB.96.115437}, in a nonlinear interacting system with level energies depending on the occupation of the levels \cite{PhysRevA.61.023402,PhysRevA.61.033603,PhysRevA.66.023404}, and in a time-dependent sweeping scheme \cite{PhysRevB.66.174438,PhysRevA.98.022102}.
All of the above studies focus on Hermitian systems that are assumed to be conservative, obey time-reversal symmetry, and obviously exhibit real-valued eigenvalues. However, in many situations, nonconservative elements arise in various forms, so that non-Hermitian physics have recently attracted considerable attention \cite{PhysRevLett.80.5243,RevModPhys.88.035002,Feng2017,Mirieaar7709,El-Ganainy2018,doi:10.1126/science.aar7709,Bouganne2020,doi:10.1080/00018732.2021.1876991,RevModPhys.93.015005}. Recently, the extensions of the LZT to non-Hermitian systems have  also been studied by considering  level decay and dephasing effect \cite{PhysRevB.36.2770,PhysRevA.46.4110,PhysRevA.90.032116,PhysRevA.104.013111}.

In the present paper, we attempt to investigate the LZT in a nonreciprocal self-interacting two-level system, in which the non-Hermiticity is induced by the nonreciprocal coupling between the levels  in contrast to  the  level decay and dephasing effects \cite{PhysRevA.46.4110,PhysRevA.90.032116,PhysRevA.104.013111}. The nonreciprocity of state transitions can be harnessed to engineer an effective non-Hermitian Hamiltonian \cite{Sounas2017,doi:10.1080/00018732.2021.1876991,RevModPhys.93.015005,Fruchart2021} that maintains the time-reversal symmetry but violates the parity symmetry. The Bogoliubov-de Gennes equation that describes the dynamics of noncondensed atoms in a Bose-Einstein condensate \cite{PhysRevLett.79.3553,PhysRevA.57.3008,PhysRevA.73.013601} has this type of symmetry. This approach has been also used to realize a non-Hermitian Su-Schrieffer-Heeger model with asymmetric intra-unit-cell hopping amplitudes \cite{PhysRevB.22.2099}. Nonreciprocity plays an important role not only in fundamental studies such as investigations of topological photonics \cite{Wang2009,doi:10.1063/1.3293411} and chiral quantum optics \cite{Lodahl2017} but also in applied research such as optical communication and information processing \cite{PhysRevLett.109.013603,Kim2015,PhysRevLett.120.060601,PhysRevApplied.9.044031}. Recently, tunable nonreciprocal hopping has been realized in cold atoms in optical lattices \cite{PhysRevX.8.031079,doi:10.1080/00018732.2019.1594094,doi:10.1080/00018732.2021.1876991} and in synthetic momentum lattices \cite{PhysRevLett.124.070402}.

Additionally, quantum adiabatic evolution is an important concept in quantum mechanics \cite{Born1927,Born1928} that has been widely applied in the preparation and control of quantum states \cite{quant-ph/0001106,doi:10.1126/science.1057726,RevModPhys.90.015002,RevModPhys.91.045001}. According to the adiabatic theorem of quantum mechanics, an initial nondegenerate eigenstate remains an instantaneous eigenstate if a Hamiltonian is changed sufficiently slowly compared to the level spacings.
It predicts a zero quantum transition between the energy levels in the adiabatic limit of the Hamiltonian's change.
 Adiabatic theory has been extended to quantum systems with nonlinear interactions \cite{PhysRevLett.90.170404, PhysRevA.81.052112}, showing many applications \cite{PhysRevLett.94.140402,PhysRevLett.98.050406,Liu2018}. Recently, in the context of non-Hermitian systems, several studies on possible new phenomena of adiabatic evolution have been carried out \cite{PhysRevA.84.023415,RevModPhys.91.045001,PhysRevA.99.032121,PhysRevA.105.013714}. In this paper, we address this issue by investigating an extended LZT model with nonreciprocal coupling, emphasizing the impact of the interplay of nonreciprocity and nonlinear interactions on the adiabatic evolution. Interestingly, we find that,
 the emergence of exceptional point (EP) singularities dramatically alters the tunneling process,
 leading to the breakdown of adiabaticity, i.e., the  emergence  of nonzero  tunneling probabilities even in the adiabatic limit of the energy bias change.
 Our results are of significance in the fields of quantum optics and quantum transport. They also have potential application in quantum devices for quantum information processing.

The paper is organized as follows. In section~\ref{SecMod}, we introduce the physical model of nonreciprocal LZT and dervice its corresponding classical Josephson Hamiltonian. In section~\ref{SecLinear}, we study the nonreciprocal LZT in the absence of nonlinear interactions. In section~\ref{SecNLinear}, in the adiabatic limit, we study the nonreciprocal LZT in the presence of nonlinear interactions. In section~\ref{SecDiscussions}, we present some remarks on the normalization issues of  the non-Hermitian systems. A summary is presented in section~\ref{SecCon}.

\section{Physical model and Josephson Hamiltonian}\label{SecMod}
\subsection{Nonreciprocal Landau-Zener model}
%%%%%%%%%%%%%%%%%%%%%%%%%%%%%%%%%%%%%%%%%%%%%%%%%%%%%%%%%%%%%%%%%%%%%
\begin{figure}[tb]
\centering
\includegraphics[width=\columnwidth]{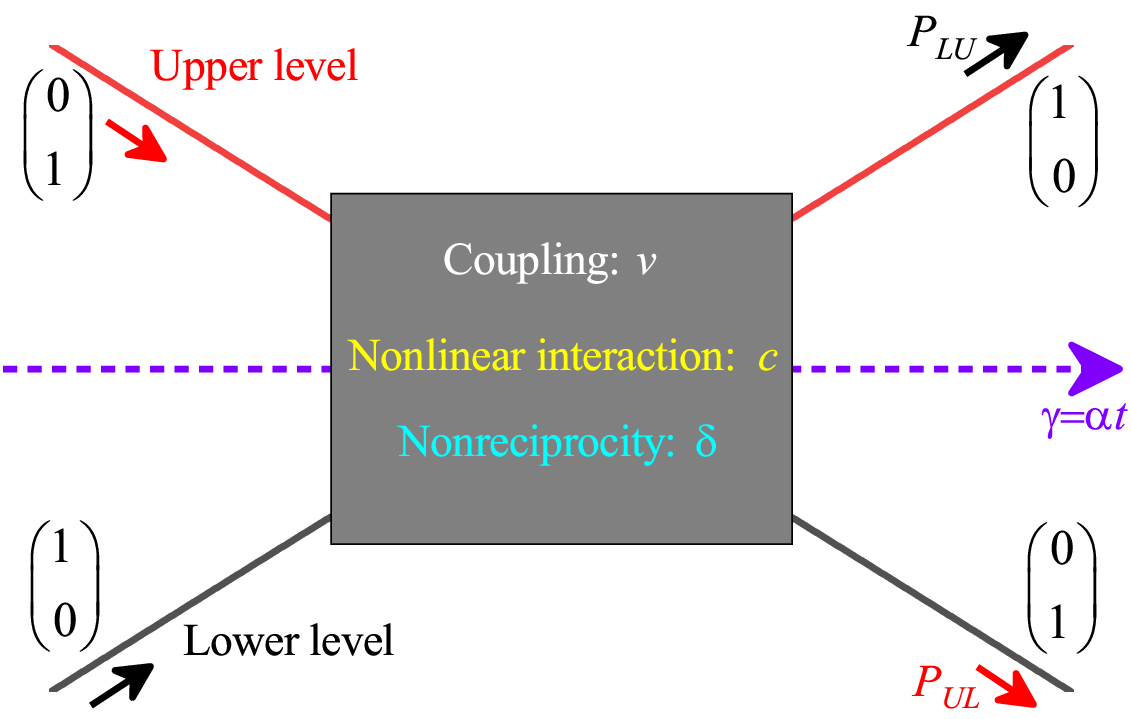}
\caption{(color online) Schematic illustration of nonreciprocal LZT based on a two-level system with nonlinear interactions. The two solid curves (red and black) represent the adiabatic energy levels (upper and lower levels, respectively). The shadow box represents the region in which the two levels closely approach each other and quantum tunneling emerges due to the complicated interplay between nonreciprocal coupling and nonlinear interactions. The level bias varies linearly in time as $\gamma=\alpha t$ ($\alpha$ is the so-called sweeping rate). Assuming that the system is initially on the lower (upper) energy level, the transition probability from the lower (upper) to the upper (lower) level during the sweeping process is denoted by $P_{\rm LU}$ ($P_{\rm UL}$). $(1,~0)^{\top} (\mathrm{or}~(0,~1)^{\top})$ represents the eigenstate when $\gamma\rightarrow\pm\infty$. }
\label{fig:model}
\end{figure}
%%%%%%%%%%%%%%%%%%%%%%%%%%%%%%%%%%%%%%%%%%%%%%%%%%%%%%%%%%%%%%%%%%%%%
We consider a nonreciprocal Landau-Zener model as illustrated by Fig. \ref{fig:model}, whose Hamiltonian takes the following form:
\begin{equation}\label{eq:ham}
H(\gamma)=\left(\begin{array}{cc}
\frac{\gamma}{2}+c|a|^{2} & \frac{v}{2} \\
\frac{v}{2}(1-\delta) & -\frac{\gamma}{2}+c|b|^{2}
\end{array}\right),
\end{equation}
where ($a$, $b$) is the two-mode wavefunction, $v$ is the constant for the hopping between the two levels, $\gamma$ is the level bias, $c$ denotes the nonlinear self-interaction parameter indicating the population-dependent level energy, and $\delta>0$ is the nonreciprocity parameter that results in non-Hermiticity. Since the Hamiltonian can be scaled by dividing by $v$, for convenience, we can set $v=1$ as the energy unit hereafter.

The dimensionless Schr\"{o}dinger equation is $i \frac{\mathrm{d}}{\mathrm{d} t}\left(\begin{array}{l}a \\ b\end{array}\right)=H(\gamma)\left(\begin{array}{l}a \\ b\end{array}\right)$.
As in the standard Landau-Zener model,  the parameter of the level bias  $\gamma$ changes linearly with time as $\gamma=\alpha t$. The constant rate $\alpha$ is the sweeping rate.

As illustrated in Fig. \ref{fig:model},  we assume that the system is initially found in the lower (or upper) energy level, i.e., $(a(t\to-\infty),~b(t\to-\infty))^T=(1,~0)^T$ (or $(a(t\to-\infty),~b(t\to-\infty))^T=(0,~1)^T$). Since Hamiltonian (\ref{eq:ham}) is non-Hermitian, the distinct feature is the appearance of complex eigenvalues. Thus, the time evolution of such a non-Hermitian system is no longer unitary and the total population (i.e., $N=n_a^2+n_b^2$ with $n_a=|a|$ and $n_b=|b|$) is not a conserved quantity. Thus, the transition probability from the lower to the upper or from the upper to the lower level can be denoted by the following:
\begin{equation}\label{eq:lzp}
P_{\rm LU}\equiv \frac{n^2_a(t\to+\infty)}{N(t\to+\infty)}; \,\, P_{\rm UL}\equiv \frac{n^2_b(t\to+\infty)}{N(t\to+\infty)}.
\end{equation}

In the absence of nonreciprocity $\delta$ and nonlinear interaction $c$, Hamiltonian (\ref{eq:ham}) degenerates into a standard linear LZT two-level system. The probability of the transition between the two energy levels is represented by the formula \cite{Landau1932,Zener1932} $P_{\rm LZ}=\exp\left(-\frac{\pi v^2}{2\alpha}\right)$. In the adiabatic limit, that is, as the sweeping rate $\alpha$ tends to zero, the transition probability tends to zero, indicating that an initial quantum state can closely follow the instantaneous eigenstate and remain in the adiabatic (upper or lower) energy level.

The presence of a nonlinear interaction $c$ can dramatically alter the tunneling dynamics \cite{PhysRevA.61.023402,PhysRevA.61.033603,PhysRevA.66.023404,PhysRevA.66.063603,PhysRevLett.87.140402,PhysRevA.65.063612,
PhysRevLett.90.170404,PhysRevLett.91.230406,PhysRevA.73.063609,PhysRevLett.96.020405,PhysRevLett.102.230401,Zenesini_2008,
PhysRevLett.103.090403,Chen2011,PhysRevLett.106.155302,Zenesini_2008,PhysRevLett.103.090403,Chen2011,PhysRevLett.120.040407,
PhysRevA.99.023616,PhysRevLett.125.213401}. The most striking feature is the breakdown of adiabaticity for a large nonlinear parameter, which is intimately linked to the hysteresis phenomena and the existence of the swallowtail loops in the adiabatic energy levels \cite{PhysRevA.61.023402,PhysRevA.66.023404,PhysRevA.66.063603,PhysRevLett.90.170404,Eckel2014}. The underlying mechanism has been revealed by investigating an equivalent classical Josephson Hamiltonian, in which the nonzero adiabatic tunneling probability can be explained as a jump in the classical canonical action \cite{PhysRevA.66.023404,PhysRevLett.90.170404}.

The presence of nonreciprocity parameter $\delta$ in Hamiltonian \eqref{eq:ham} can result in nonreciprocal state transitions between two levels.
In contrast to the $\mathcal{PT}$-conserved non-Hermitian systems \cite{PhysRevLett.80.5243}, this non-Hermitian system maintains the time-reversal symmetry but violates the parity symmetry.

In the present paper, we focus on the adiabaticity in the nonreciprocal LZT by investigating the  tunneling probabilities in the adiabatic limit of the energy bias change, i.e., $P^{\rm ad}_{\rm LU}=\lim_{\alpha\to 0} P_{\rm LU}$ and $P^{\rm ad}_{\rm UL}=\lim_{\alpha\rightarrow0} P_{\rm UL}$.
According to the adiabatic theorem, when adiabaticity is maintained,  the adiabatic tunneling probabilities should be zero. Nevertheless, the emergence of EPs can cause the breakdown of the adiabaticity  and lead to nonzero  adiabatic tunneling probability. Our theoretical analysis is facilitated  by  the following Lagrangian representation that help us to derive a classical Josephson Hamiltonian.

\subsection{Lagrangian and Classical Josephson Hamiltonian}
The  Schr\"{o}dinger equation that governs  the nonreciprocal LZT can be derived from the time-dependent variational principle, where the action is defined as the time integral of a Lagrangian and is extremized with respect to the two-mode wavefunction $\psi(\gamma=\alpha t)=(a(t),~b(t))^T$. Here, the Lagrangian is found to be as follows:
\begin{equation}\label{eq:Lagrangian1}
L=\left\langle\psi\left|i \sigma \frac{\mathrm{d}}{\mathrm{d} t}\right| \psi\right\rangle-\left\langle\psi\left|\sigma H_{0}\right| \psi\right\rangle-\frac{c}{4}\left\langle\psi\left|\sigma\left(\sigma_{z}\rho\sigma_{z}+\rho\right) \right| \psi\right\rangle ,
\end{equation}
where $\sigma=\left(\begin{array}{cc} 1-\delta & 0 \\ 0 & 1 \end{array}\right)$, $H_0=\left(\begin{array}{cc} \frac{\gamma}{2} & \frac{v}{2} \\ \frac{v}{2}(1-\delta) & -\frac{\gamma}{2} \end{array}\right)$ is the linear part of original Hamiltonian \eqref{eq:ham}, $\sigma_z=\left(\begin{array}{cc} 1 & 0 \\ 0 & -1 \end{array}\right)$ is the Pauli matrix, and  $\rho=|\psi\rangle\langle\psi|$ is the density matrix. The wavefunction can be expressed in its amplitude and phase by $a=n_ae^{i\theta_a}$ and $b=n_be^{i\theta_b}$, and the Lagrangian is then cast into the following form:
\begin{equation}\label{eq:Lagrangian2}
\begin{aligned}
L =& i(1-\delta)\dot{n}_{a} n_{a}-(1-\delta) \dot{\theta}_{a} n_{a} n_{a}+i \dot{n}_{b} n_{b}-\dot{\theta}_{b} n_{b} n_{b} \\
&-\frac{1}{2} \gamma\left[(1-\delta) n_{a} n_{a}-n_{b} n_{b}\right]-v(1-\delta) n_{a} n_{b} \cos \left(\theta_{b}-\theta_{a}\right) \\
&-\frac{c}{2}\left((1-\delta) n_{a}^{4}+n_{b}^{4}\right) .
\end{aligned}
\end{equation}
One can naturally obtain a classical Hamiltonian from the Lagrangian \eqref{eq:Lagrangian2} using the relationship $H^c=\left\langle\psi\left|i \sigma \frac{\mathrm{d}}{\mathrm{d} t}\right| \psi\right\rangle-L$.

We then introduce the following four canonical variables,
\begin{subequations} \label{eq:CanonicalVariables}
\begin{align}
&\theta_+=\theta_1+\theta_2 \;,\\
&\theta_-=\theta_1-\theta_2 \;,\\
&\zeta_+=n_a^2+\frac{1}{1-\delta}n_b^2 \;,\\
&\zeta_-=n_a^2-\frac{1}{1-\delta}n_b^2 \;.
\end{align}
\end{subequations}
In terms of $\theta_+$, $\theta_-$, $\zeta_+$, and $\zeta_-$, the classical Hamiltonian can be expressed as follows:
\begin{equation}\label{eq:Hamiltonian1}
\begin{aligned}
H^c=&\gamma \zeta_{-}+v \sqrt{(1-\delta)\left(\zeta_{+}^{2}-\zeta_{-}^{2}\right)} \cos \theta_{-} \\
&+\frac{c (2-\delta)}{4}\left(\zeta_{+}^{2}+\zeta_{-}^{2}\right)+\frac{c\delta}{2} \zeta_{+} \zeta_{-} \;.
\end{aligned}
\end{equation}
We note that the above classical Hamiltonian \eqref{eq:Hamiltonian1}  can be reduced to either the famous Josephson Hamiltonian \cite{Anderson1963,PhysRevB.40.2158,PhysRevLett.79.4950} when $\delta=c=0$, or to its nonlinear extension  when $\delta=0$ and $c \ne 0$ \cite{PhysRevA.66.023404,Liu2018}.

From the Hamilton's canonical equations $\dot{\theta}_{-}=-\partial H^c / \partial \zeta_-$, $\dot{\theta}_{+}=-\partial H^c / \partial \zeta_+$, $\dot{\zeta}_-=\partial H^c / \partial \theta_-$, and $\dot{\zeta}_+=\partial H^c / \partial \theta_+$, one can readily obtain the following equations of motion,
\begin{subequations} \label{eq:dyn1}
\begin{align}
&\dot{\zeta}_{+}=0,\\
&\dot{\zeta}_{-}=-v \sqrt{(1-\delta)\left(\zeta_{+}^{2}-\zeta_{-}^{2}\right)} \sin \theta_{-}, \\
&\dot{\theta}_{+}=-\frac{c}{2}\delta \zeta_{-}-\frac{c}{2}(2-\delta) \zeta_{+}- \frac{(1-\delta) v\zeta_{+}\cos \theta_{-}}{\sqrt{(1-\delta)\left(\zeta_{+}^{2}-\zeta_{-}^{2}\right)}} , \\
&\dot{\theta}_{-}=-\gamma-\frac{c}{2} \delta\zeta_{+}-\frac{c}{2}(2-\delta) \zeta_{-}+\frac{(1-\delta) v \zeta_{-} \cos \theta_{-}}{\sqrt{(1-\delta)\left(\zeta_{+}^{2}-\zeta_{-}^{2}\right)}}.
\end{align}
\end{subequations}

Clearly, $\zeta_+$ is an invariant that is determined by the initial state. Nevertheless, when the parameters change adiabatically, an initial state  may encounter EPs; in this situation, we find that the  singularity of the EPs may cause a sudden change in the invariant as will be shown in the next section.

In terms of $\zeta_+$ and $\zeta_-$, the total population $N=n_a^2+n_b^2=\frac{2-\delta}{2}\zeta_{+}+\frac{\delta}{2}\zeta_{-}$. From Eq. \eqref{eq:dyn1}, we can also deduce  the following differential equation:
\begin{equation} \label{eq:dynN}
\dot{N}=-\frac{\delta v}{2}\sqrt{(1-\delta)\left(\zeta_{+}^{2}-\zeta_{-}^{2}\right)} \sin \theta_{-}    \;.
\end{equation}

\section{Nonreciprocal LZT  in the absence of nonlinear interactions ($c=0$)}\label{SecLinear}
\subsection{Energy level and nonreciprocal LZT}
%%%%%%%%%%%%%%%%%%%%%%%%%%%%%%%%%%%%%%%%%%%%%%%%%%%%%%%%%%%%%%%%%%%%%
\begin{figure}[tb]
\centering
\includegraphics[width=\columnwidth]{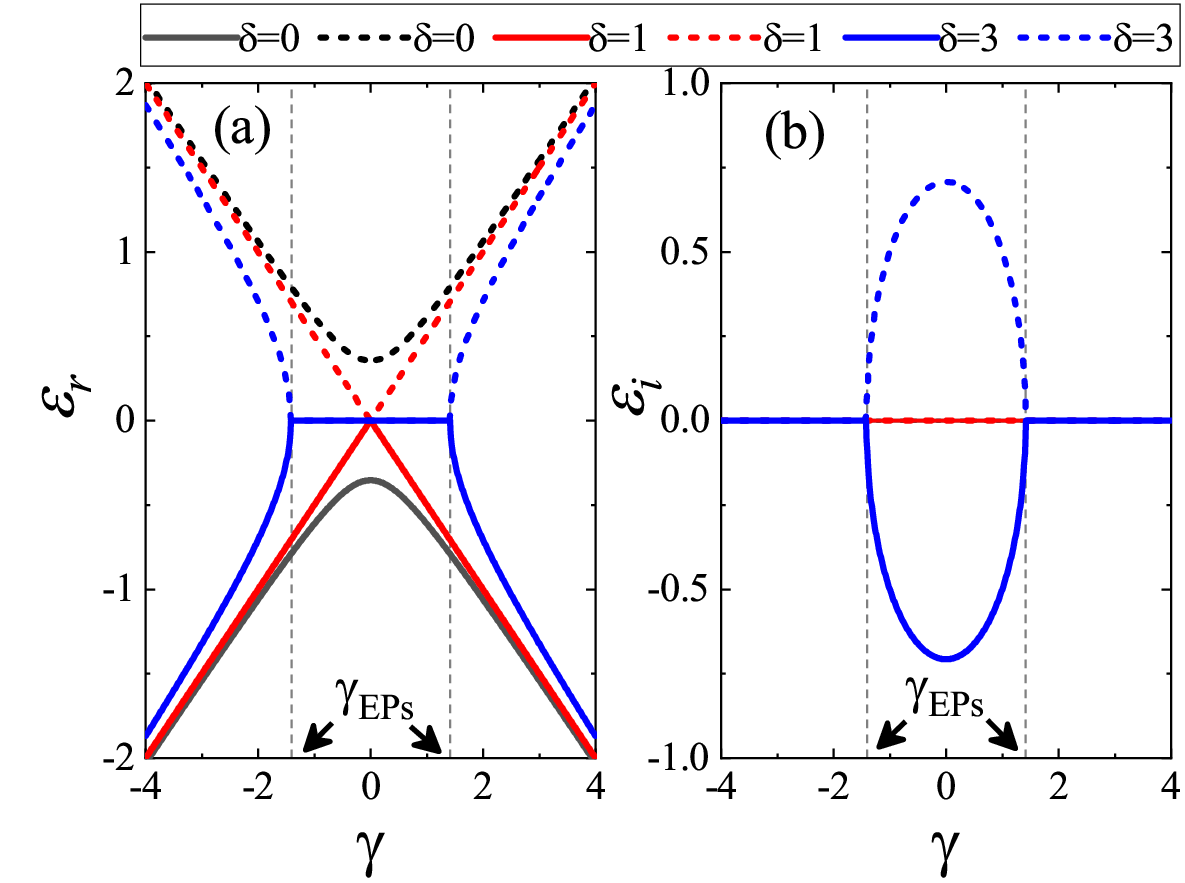}
\caption{(color online) Adiabatic levels $\varepsilon$ versus bias $\gamma$ for different nonreciprocity parameters $\delta=0$ (black curve), $1$ (red curve), and $3$ (blue curve). (a) and (b) present the real and imaginary parts of the adiabatic levels, respectively. For $\delta\leq1$, the imaginary parts of energy are always $0$, and they overlap in (b). For $\delta>1$, with increasing $\gamma$, EPs occur at $\pm\gamma_{\rm EPs}$, and the corresponding eigenvalues coalesce. In both (a) and (b), the dashed magenta lines mark the location of EPs. }
\label{fig:linearlevel}
\end{figure}
%%%%%%%%%%%%%%%%%%%%%%%%%%%%%%%%%%%%%%%%%%%%%%%%%%%%%%%%%%%%%%%%%%%%%
When $c=0$, the eigenvalues and eigenstates of Hamiltonian \eqref{eq:ham} can be readily obtained. The adiabatic levels are given by the following:
\begin{equation}\label{eq:eigenenergies}
\varepsilon_{\pm}=\pm\frac{1}{2}\sqrt{\gamma^2+v^2(1-\delta)}\;.
\end{equation}
The corresponding eigenstates are given by the following:
\begin{equation}\label{eq:eigenstates}
\begin{aligned}
&|\varepsilon_{-}\rangle=C_{1}\left(\frac{\gamma -\sqrt{\gamma ^2+v^2(1-\delta )}}{v(1-\delta ) },1\right)^T\;, \\
&|\varepsilon_{+}\rangle=C_{2}\left(\frac{\gamma +\sqrt{\gamma ^2+v^2(1-\delta )}}{v(1-\delta ) },1\right)^T\;.
\end{aligned}
\end{equation}
Here, $C_1$ and $C_2$ are normalization coefficients.

EPs at which two eigenvalues and their corresponding eigenvectors coalesce and become degenerate are a distinct feature in non-Hermitian quantum systems.
For nonreciprocity parameter $\delta \geq 1$, the conditions for EPs can be easily obtained, that is, $\gamma_{\rm EPs}=\pm v\sqrt{\delta-1}$.
 The eigenvalues are real when $|\gamma|>|\gamma_{\rm EPs}|$; otherwise, they are imaginary. The corresponding eigenvectors at EPs coalesce as described by the following:
\begin{equation}\label{eq:EPeigenstates}
|\varepsilon_{\rm EP}\rangle_{\pm}=\left(\mp\frac{1}{\sqrt{\delta}},~\frac{\sqrt{\delta-1}}{\sqrt{\delta}}\right)^T\;,
\end{equation}
where the sign $\mp$ is for the $\pm\gamma_{\rm EPs}$ EPs, respectively.
For nonreciprocity parameter $\delta < 1$, there are no EPs, and the eigenvalues are always real.

The level bias $\gamma$-dependence of the energy levels is shown in Fig. \ref{fig:linearlevel}. When $\delta<1$, the imaginary parts of the eigenvalues are always zero, and the real parts of the eigenvalues show an avoided level crossing at $\gamma=0$. For $\delta=1$, the eigenvalues show a level crossing at $\gamma=0$. For $\delta>1$, with increasing $\gamma$, EPs occur at $\gamma_{\rm EPs}$, and the corresponding eigenvalues coalesce. Between two EPs, the real parts of the eigenvalues are zero, whereas the imaginary parts of the eigenvalues are nonzero.

Note that the corresponding eigenstates are not orthogonal to each other. For $\gamma\rightarrow\pm\infty$, we have $\varepsilon\rightarrow\pm|\gamma|/2$. For instance, for the lower level, we have $(a,~b)^T\rightarrow(1,~0)^T$ at $\gamma\rightarrow-\infty$ and $(a,~b)^T\rightarrow(0,~1)^T$ at $\gamma\rightarrow+\infty$.

%%%%%%%%%%%%%%%%%%%%%%%%%%%%%%%%%%%%%%%%%%%%%%%%%%%%%%%%%%%%%%%%%%%%%
\begin{figure}[tb]
\centering
\includegraphics[width=\columnwidth]{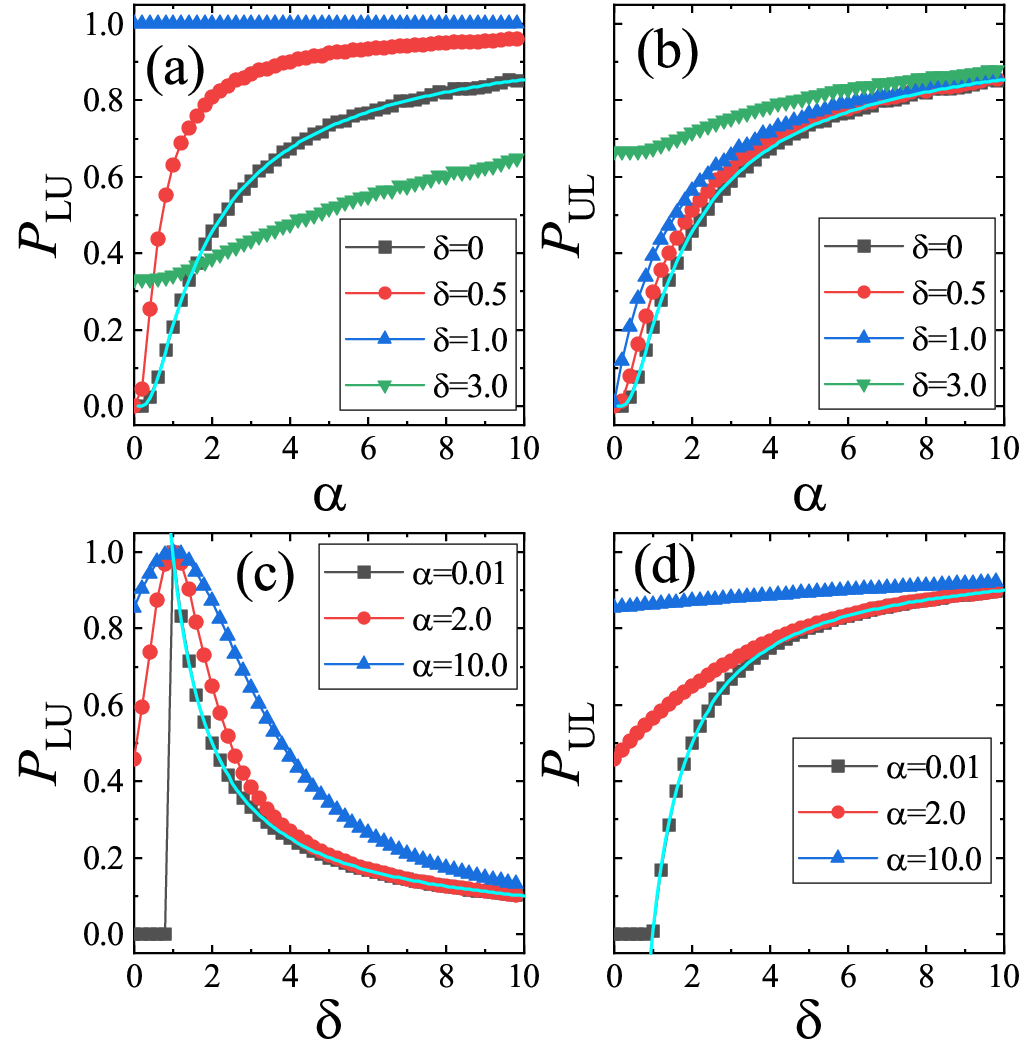}
\caption{(color online) Nonreciprocal LZT probability. (a) and (b) present the tunneling probabilities $P_{\rm LU}$ and $P_{\rm UL}$, respectively, as a function of sweeping rate $\alpha$ for different nonreciprocity parameters $\delta=0$, $0.5$, $1$, and $3$. (c) and (d) present the tunneling probabilities $P_{\rm LU}$ and $P_{\rm UL}$, respectively, as a function of nonreciprocity parameter $\delta$ for different sweeping rates $\alpha=0.01$, $2$, and $10$. In both (a) and (b), the cyan curve indicates the standard LZT probability. The cyan curves in (c) and (d) indicate the functions with forms of $1/\delta$ and $(\delta-1)/\delta$, respectively. }
\label{fig:fig3}
\end{figure}
%%%%%%%%%%%%%%%%%%%%%%%%%%%%%%%%%%%%%%%%%%%%%%%%%%%%%%%%%%%%%%%%%%%%%
We now focus on the nonreciprocal LZT. The numerical results of the tunneling probability for different sweeping rates $\alpha$ and nonreciprocity parameters $\delta$ are shown in Fig. \ref{fig:fig3}. Figures \ref{fig:fig3}(a) and (b) show the tunneling probabilities $P_{\rm LU}$ and $P_{\rm UL}$, respectively, as a function of the sweeping rate $\alpha$ for different nonreciprocity parameters $\delta=0$, $0.5$, $1$, and $3$. For $\delta=0$, both the tunneling probabilities $P_{\rm LU}$ and $P_{\rm UL}$ are consistent with the standard formula of the LZT probability. For $\delta\neq0$, nonreciprocity greatly changes the LZT, resulting in breaking of the symmetry in the tunneling between the two levels, i.e., $P_{\rm LU}\neq P_{\rm UL}$. For $\delta=0.5$, the tunneling probabilities $P_{\rm LU}$ and $P_{\rm UL}$ are both exponentially small as the value of $\alpha$ approaches zero. For $\delta=1$, with the increase in $\alpha$, the tunneling probability $P_{\rm LU}$ remains $1$, whereas the tunneling probability $P_{\rm UL}$ is exponentially small as the value of $\alpha$ approaches zero. For $\delta=3$, the tunneling probabilities $P_{\rm LU}$ and $P_{\rm UL}$ both decrease as $\alpha$ decreases. For subcritical values of the nonreciprocity parameter $\delta<1$, the tunneling probability still exponentially vanishes with $\alpha$. Most strikingly, the tunneling probabilities $P_{\rm LU}$ and $P_{\rm UL}$ for $\delta>1$ are both not zero in the adiabatic limit $\alpha\rightarrow 0$, while they are zero for $\delta<1$. This shows that $\delta=1$ is the critical point for the adiabatic time evolution.

Figures \ref{fig:fig3}(c) and (d) more clearly depict the tunneling probabilities $P_{\rm LU}$ and $P_{\rm UL}$, respectively, as a function of the nonreciprocity parameter $\delta$ for different sweeping rates $\alpha=0.01$, $2$, and $10$. In the adiabatic limit, i.e., $\alpha=0.01$, the tunneling probabilities $P_{\rm LU}$ and $P_{\rm UL}$ both remain zero when $\delta<1$. At $\delta=1$, $P_{\rm LU}$ quickly reaches 1, whereas $P_{\rm LU}$ remains 0. For $\delta>1$, with increasing $\delta$, $P_{\rm LU}$ decreases as a function with the form of $1/\delta$, and $P_{\rm UL}$ increases as a function with the form of $(\delta-1)/\delta$. In the nonadiabatic cases, i.e., $\alpha=2,~10$, as $\delta$ increases from $0$ to $1$, $P_{\rm LU}$ increases continuously. When $\delta=1$, $P_{\rm LU}$ reaches the saturation value of $1$ regardless of $\alpha$; then, as $\delta$ continues to increase, $P_{\rm LU}$ keeps decreasing. When $\delta$ reaches a sufficiently large value, $P_{\rm LU}$ decreases to $0$. In the nonadiabatic cases, $P_{\rm UL}$ increases as $\delta$ increases.

From the above results, we find that even in the adiabatic limit $\alpha\rightarrow 0$, the tunneling probability can take nonzero values, depending solely on the nonreciprocity parameter in the forms of $1/\delta$ and $(\delta-1)/\delta$, as indicated by the cyan curves in Figs. \ref{fig:fig3}(c) and \ref{fig:fig3}(d).

\subsection{Nonreciprocity breaks adiabaticity }
%%%%%%%%%%%%%%%%%%%%%%%%%%%%%%%%%%%%%%%%%%%%%%%%%%%%%%%%%%%%%%%%%%%%%
\begin{figure}[tb]
\centering
\includegraphics[width=\columnwidth]{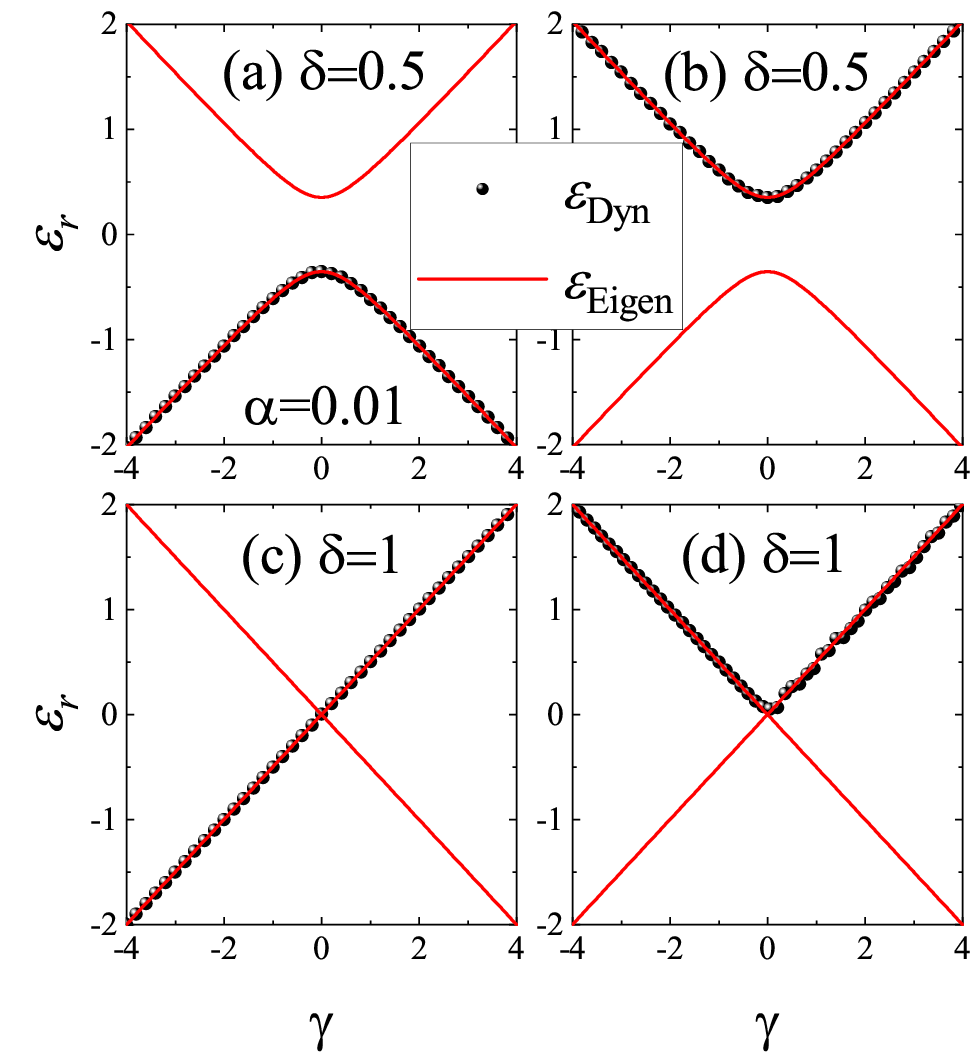}
\caption{(color online) Comparison of the dynamical levels $\varepsilon_{\mathrm{Dyn}}$ (circles) and the adiabatic levels $\varepsilon_{\mathrm{Eigen}}$ (solid lines) for $\delta\leq1$. (a) and (b) present the results for $\delta=0.5$, and (c) and (d) present the results for $\delta=1$. For dynamical evolution, we take the adiabatic limit of $\alpha=0.01$ as an example. }
\label{fig:fig4}
\end{figure}
%%%%%%%%%%%%%%%%%%%%%%%%%%%%%%%%%%%%%%%%%%%%%%%%%%%%%%%%%%%%%%%%%%%%%

To gain insight into the adiabaticity in nonreciprocal LZT, we calculate the dynamical energy in the temporal evolution \cite{PhysRevA.61.023402,PhysRevLett.125.213401}. For the wavefunction $\psi(\gamma=\alpha t)=(a(t),~b(t))^T$ of the Schr\"{o}dinger equation with the Hamiltonian (\ref{eq:ham}), the  dynamical energy can be obtained by calculating the energy expectation values of the dynamic states:
\begin{equation}\label{eq:dynE}
\varepsilon_{\mathrm{Dyn}}(\gamma)=\frac{\langle\psi(\gamma=\alpha t)|H(\gamma)|\psi(\gamma=\alpha t)\rangle}{\langle\psi(\gamma=\alpha t)|\psi(\gamma=\alpha t)\rangle}.
\end{equation}
According to the adiabatic theorem, when adiabaticity is maintained, the dynamical energy level  at a very small sweeping rate will  perfectly follow the adiabatic energy  level. A situation where the dynamical energy cannot  follow the adiabatic level even for a very small sweeping rate indicates the  breakdown of adiabaticity and  the system manifests a nonzero adiabatic tunneling probability  \cite{PhysRevA.61.023402,PhysRevLett.125.213401}.
A comparison of the dynamical levels $\varepsilon_{\mathrm{Dyn}}$ and the adiabatic levels $\varepsilon_{\mathrm{Eigen}}$ is shown in Fig. \ref{fig:fig4}. Figures \ref{fig:fig4}(a) and (b) show the results for $\delta=0.5$, and (c) and (d) show those for $\delta=1$. The initial state is prepared in the lower level (i.e., $(a,~b)^T=(1,~0)^T$) for Figs. \ref{fig:fig4}(a) and (c) and in the upper level (i.e., $(a,~b)^T=(0,~1)^T$) for Figs. \ref{fig:fig4}(b) and (d). The results are shown in Fig. \ref{fig:fig4} for the adiabatic limit with $\alpha=0.01$. In Figs. \ref{fig:fig4}(a) and (b), we observe an excellent match between the dynamical levels and the adiabatic levels, which is similar to the reciprocal  case of $\delta=0$ \cite{PhysRevA.61.023402}. Our further calculations demonstrate that for  $\delta<1$, the quantum state can closely follow the corresponding eigenstates in the adiabatic evolution, indicating that adiabaticity is maintained.

%%%%%%%%%%%%%%%%%%%%%%%%%%%%%%%%%%%%%%%%%%%%%%%%%%%%%%%%%%%%%%%%%%%%%
\begin{figure}[tb]
\centering
\includegraphics[width=\columnwidth]{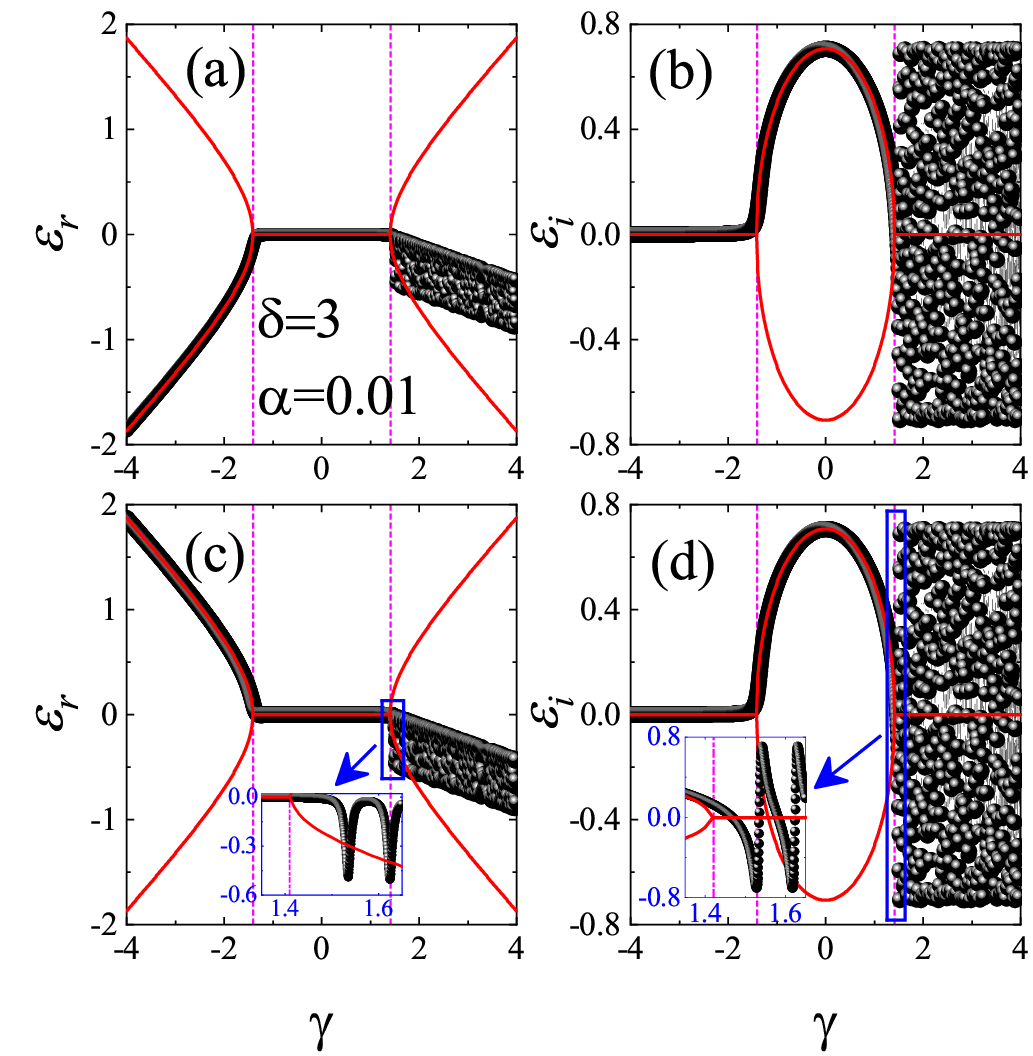}
\caption{(color online) Comparison of the dynamical levels $\varepsilon_{\mathrm{Dyn}}$ (circles) and the adiabatic levels $\varepsilon_{\mathrm{Eigen}}$ (solid lines) for $\delta>1$. In this case, the EPs occur at $\gamma=\gamma_{\rm EPs}$, which are marked by two magenta dotted lines. (a) and (c) present the real parts of the energy levels, and (b) and (d) present the imaginary parts of the energy levels. For dynamical evolution, the system is initially prepared in the lower adiabatic levels for (a) and (b), i.e., $(a,~b)^T=(1,~0)^T$, whereas the system is initially prepared in the upper adiabatic levels for (c) and (d), i.e., $(a,~b)^T=(0,~1)^T$. Here, we take $\delta=3$ and the adiabatic limit as $\alpha=0.01$ as an example. }
\label{fig:fig5}
\end{figure}
%%%%%%%%%%%%%%%%%%%%%%%%%%%%%%%%%%%%%%%%%%%%%%%%%%%%%%%%%%%%%%%%%%%%%

At the critical value of $\delta=1$, the upper level and lower level cross at $\gamma=0$. We calculate the dynamical energy levels, shown in Figs. \ref{fig:fig4} (c) and (d).
The results show that the initial state starting from both the upper level and lower level can finally evolve into the upper level.
When $\delta=1$, the adiabatic levels of Hamiltonian \eqref{eq:ham} are given by $\varepsilon_{\pm}|_{\delta=1}=\pm\frac{1}{2}\gamma$, and the corresponding adiabatic states are given by $|\varepsilon_{-}\rangle_{\delta=1}=(-v/2\gamma,~1)^T$ and $|\varepsilon_{+}\rangle_{\delta=1}=(1,~0)^T$.
For $\gamma\rightarrow0$, both eigenvalues $\varepsilon_{\pm}|_{\delta=1}\rightarrow0$ and the normalized  eigenvectors coalesce as $|\varepsilon_{\pm}\rangle_{\delta=1}\rightarrow(1,~0)^T$, indicating the appearance of an EP.
Because  the term of $v(1-\delta)/2$ in Hamiltonian  (1)  vanishes when $\delta=1$,
 the  Schr\"{o}dinger equation can be solved analytically as $a(t)=a_0 e^{-\frac{i \alpha t^2}{4}} -\frac{\left(\frac{1}{4}+\frac{i}{4}\right) \sqrt{\pi } b_0 v e^{-\frac{1}{4} i \alpha  t^2} \text{erfi}\left(\left(\frac{1}{2}+\frac{i}{2}\right) \sqrt{\alpha } t\right)}{\sqrt{\alpha }}$, and $b(t)=b_0e^{\frac{i \alpha t^2}{4}}$, where the symbol $\text{erfi}$ represents the imaginary error function, and coefficient vector  $(a_0, b_0)^T$ is given  by the normalized eigenvector $(1,~0)^T$ at EP. Then, we have  $a(t)=e^{-\frac{i \alpha t^2}{4}}$ and $b(t)=0$ for $t > 0$. Thus,  the normalized wavefunction is $(\tilde{a}(t),~\tilde{b}(t))^T=(1,~0)^T$.  This indicates that the initial state starting from either the upper level or the lower level will finally evolve into the upper level.

For $\delta>1$, EPs occur at $\gamma=\pm\gamma_{\rm EPs}$. Therefore, when $\gamma$ changes from $-\infty$ to $+\infty$, the system will pass through the EPs; here, the adiabaticity may break. A comparison of the dynamical levels $\varepsilon_{\mathrm{Dyn}}$ and the adiabatic levels $\varepsilon_{\mathrm{Eigen}}$ for $\delta=3$ is shown in Fig. \ref{fig:fig5}. Figures \ref{fig:fig5}(a) and (b) show the real part and imaginary part of the energy levels.
It is observed that in the range of $\gamma<\gamma_{\rm EPs}$, the initial state prepared in either the upper or lower level can perfectly follow the real part of the adiabatic energy level.
However, in the range from $-\gamma_{\rm EPs}$ to $+\gamma_{\rm EPs}$, both states evolve following the imaginary part of the upper level. This is due to the so-called skin effect \cite{RevModPhys.93.015005}, in which in the non-Hermitian system, the eigenvalue with a positive imaginary part will dominate the dynamical evolution.
More interestingly, when $\gamma>+\gamma_{\rm EPs}$, the quantum states follow neither the upper level nor the lower level, clearly indicating the breakdown of adiabaticity.

%%%%%%%%%%%%%%%%%%%%%%%%%%%%%%%%%%%%%%%%%%%%%%%%%%%%%%%%%%%%%%%%%%%%%
\begin{figure}[tb]
\centering
\includegraphics[width=\columnwidth]{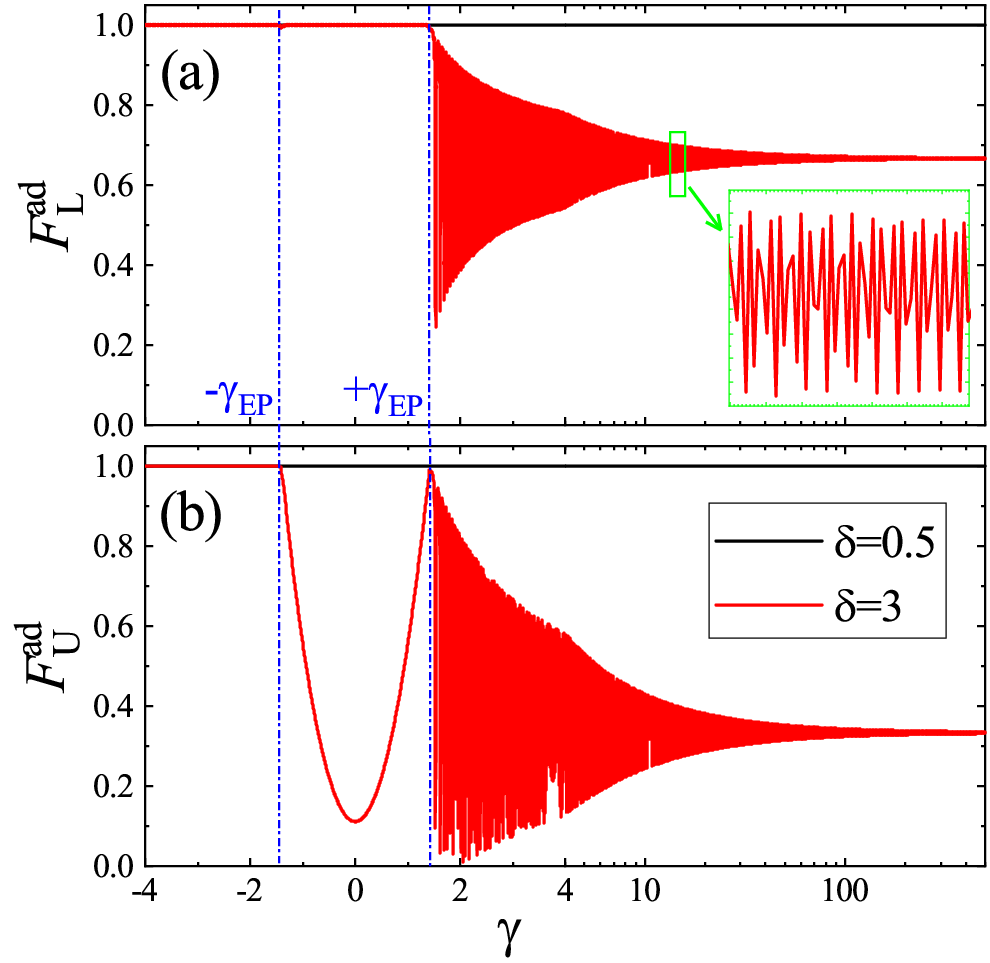}
\caption{(color online) The adiabatic fidelities $F_{\rm U}^{\rm ad}$ and $F_{\rm L}^{\rm ad}$ of a normalized dynamic state $|\psi(\gamma=\alpha t)\rangle$ versus the level bias $\gamma=\alpha t$ in the adiabatic limit (i.e., $\alpha=0.01$). For dynamical evolution, the system is initially prepared in the lower and upper adiabatic levels for (a) and (b), respectively. In both (a) and (b), we adopt logarithmic coordinates in the range of $\gamma>4$ to clearly exhibit the convergence behavior of adiabatic fidelities. For the larger enough $\gamma$, the adiabatic fidelity finally converges to a certain value, i.e., $F_{\rm L}^{\rm ad}(\gamma\to+\infty)=1-P_{\rm LU}^{\rm ad}$ in (a), and $F_{\rm U}^{\rm ad}(\gamma\to+\infty)=1-P_{\rm UL}^{\rm ad}$ in (b), respectively. }
\label{fig:fig501}
\end{figure}
%%%%%%%%%%%%%%%%%%%%%%%%%%%%%%%%%%%%%%%%%%%%%%%%%%%%%%%%%%%%%%%%%%%%%

The adiabatic fidelities can be introduced to measure how close a normalized quantum state $|\psi(\gamma=\alpha t)\rangle$   is to the  adiabatic eigenstate $|\varepsilon_{\pm}\rangle$ \cite{Liu2018}, i. e.,  $F_{\rm U}^{\rm ad}(\gamma)=\lim_{\alpha\to 0}\frac{\left|\langle\psi(\gamma=\alpha t)|\varepsilon_{+}(\gamma)\rangle\right|^2}{\langle\psi(\gamma=\alpha t)|\psi(\gamma=\alpha t)\rangle}$ and $F_{\rm L}^{\rm ad}(\gamma)=\lim_{\alpha\to 0}\frac{\left|\langle\psi(\gamma=\alpha t)|\varepsilon_{-}(\gamma)\rangle\right|^2}{\langle\psi(\gamma=\alpha t)|\psi(\gamma=\alpha t)\rangle}$. The convergence of the adiabatic fidelity to unit uniformly over the range $\gamma\in(-\infty,~+\infty)$ indicates the preservation of the adiabaticity in the quantum evolution.

In Fig. \ref{fig:fig501}, we show the adiabatic fidelities $F_{\rm U}^{\rm ad}$ and $F_{\rm L}^{\rm ad}$ of the dynamic state $|\psi(\gamma=\alpha t)\rangle$ versus the level bias $\gamma=\alpha t$ for the very small sweeping rate of  $\alpha=0.01$. In Fig. \ref{fig:fig501}(a), we suppose that the initial state is the lower adiabatic state, i.e., $(a(\gamma\to-\infty),~b(\gamma\to-\infty))^T=(1,~0)^T$. For the small nonreciprocity parameter ($\delta<1$), i.e., $\delta=0.5$, the adiabatic fidelity remains unitary over the range $\gamma\in(-\infty,~+\infty)$, indicating the preservation of the adiabaticity. For the larger nonreciprocity parameter ($\delta>1$), i.e., $\delta=3$, the adiabatic fidelity remains unitary in the range $\gamma\in(-\infty,~+\gamma_{\rm EP})$, whereas, it becomes oscillatory in the range $\gamma>+\gamma_{\rm EP}$, indicating the breakdown of adiabaticity. For large enough $\gamma$, the adiabatic fidelity finally converges to a certain value, i.e., $F_{\rm L}^{\rm ad}(\gamma\to+\infty)=1-P_{\rm LU}^{\rm ad}$. In Fig. \ref{fig:fig501}(b), the initial state is prepared in the upper adiabatic eigenstates, i.e., $(a(\gamma\to-\infty),~b(\gamma\to-\infty))^T=(0,~1)^T$. We observed that the adiabatic fidelity remains unitary for the small nonreciprocity parameter ($\delta<1$). For the larger nonreciprocity parameter ($\delta>1$), the adiabatic fidelity remains unitary only in the range $\gamma\in(-\infty,~-\gamma_{\rm EP})$, and becomes oscillatory in the range $\gamma>-\gamma_{\rm EP}$, indicating the breakdown of adiabaticity. For large enough $\gamma$, the adiabatic fidelity finally converges to be  $F_{\rm U}^{\rm ad}(\gamma\to+\infty)=1-P_{\rm UL}^{\rm ad}$.

\subsection{Analytical deductions of the adiabatic  probabilities of $P_{\rm UL}^{\rm ad}$ and $P_{\rm LU}^{\rm ad}$}\label{SecLinearC}
In terms of $\zeta_+$ and $\zeta_-$, the formula \eqref{eq:lzp} of the  transition probability can be rewritten as,
\begin{subequations} \label{eq:pad}
\begin{align}
&P_{\rm LU}=\frac{\zeta_+(t\to+\infty)+\zeta_-(t\to+\infty)}{(2-\delta)\zeta_+(t\to+\infty)+\delta\zeta_-(t\to+\infty)}, \label{eq:pada}\\
&P_{\rm UL}=\frac{(1-\delta)\left[\zeta_+(t\to+\infty)-\zeta_-(t\to+\infty)\right]}{(2-\delta)\zeta_+(t\to+\infty)+\delta\zeta_-(t\to+\infty)} .\label{eq:padb}
\end{align}
\end{subequations}

Considering that the initial state is on lower level and according to Eq. \eqref{eq:CanonicalVariables} and explicit expressions of the eigenstates in Eq. \eqref{eq:eigenstates}, we readily obtain
\begin{subequations} \label{eq:zetaeig}
\begin{align}
&\zeta_{+}(t)=\frac{N(\gamma)}{1-\frac{\delta}{2}\left(1+\frac{\gamma}{\sqrt{\gamma^2+v^2(1-\delta)}}\right)},\label{eq:zetaeiga}\\
&\zeta_{-}(t)=\frac{2\gamma N(\gamma)}{(\delta-2)\sqrt{\gamma^2+v^2(1-\delta)}+\delta\gamma} .\label{eq:zetaeigb}
\end{align}
\end{subequations}
It should be noted that $N(\gamma)$ appears in the above formula and the total population is not always unitary in the non-Hermitian system.

In the weak nonreciprocal regime ($\delta<1$),  in the adiabatic limit, the quantum state evolves on the lower level.  When $\gamma\to+\infty$, we have $\zeta_{+}(t\to+\infty)=\frac{N(t\to+\infty)}{1-\delta}$ and  $\zeta_{-}(t\to+\infty)=-\frac{N(t\to+\infty)}{1-\delta}$. Substituting these relations into Eq. \eqref{eq:pada}, we obtain the following:
\begin{equation}
P_{\rm LU}^{\rm ad}=0~~~~~~(\delta<1)\;.
\end{equation}
Similar analysis applies to the upper level that  gives $P_{\rm UL}^{\rm ad}=0$ when $\delta<1$.

While in the strong nonreciprocal regime ($\delta>1$), the adiabatic quantum state will encounter EPs at $\gamma_{\rm EP}=-v\sqrt{\delta-1}$, where the denominator of the expression of $\zeta_{+}$ in Eq. \eqref{eq:zetaeiga} diverges. This singularity  allows
$\zeta_{+}$ to suddenly become zero at $\gamma_{\rm EP}$.
After EP, with increasing $\gamma$, we find that $\zeta_{+}$ remains zero while the total population $N$ will grow rapidly. This is somewhat similar to the so-called Pancharatnam phenomenon \cite{Berry1994}.
The Pancharatnam phenomenon concerns the time development of a parameter-dependent non-Hermitian two-level system and indicates that the quantum state will eventually evolve into the eigenstate of EP when the system passes through the EP slowly.
Therefore, when $\gamma\to+\infty$, we have $\zeta_{+}(t\to+\infty)=\zeta_{+}(\gamma_{\rm EP})=0$. Substituting this relation into Eq. \eqref{eq:pada}, we  immediately obtain $P_{\rm LU}^{\rm ad}=\frac{1}{\delta}$.
Similar analysis applies to the upper level. We finally obtain
the adiabatic LZT probability for $\delta>1$:
\begin{equation} \label{eq:adiabaticP}
P_{\rm LU}^{\rm ad}=\frac{1}{\delta} \;;~~~~ P_{\rm UL}^{\rm ad}=\frac{\delta-1}{\delta}~~~~~~(\delta>1)\;.
\end{equation}
The above analytic expressions are in full agreement with the numerical results, as shown in Figs. \ref{fig:fig3}(c) and (d).

%%%%%%%%%%%%%%%%%%%%%%%%%%%%%%%%%%%%%%%%%%%%%%%%%%%%%%%%%%%%%%%%%%%%%
\begin{figure}[tb]
\centering
\includegraphics[width=\columnwidth]{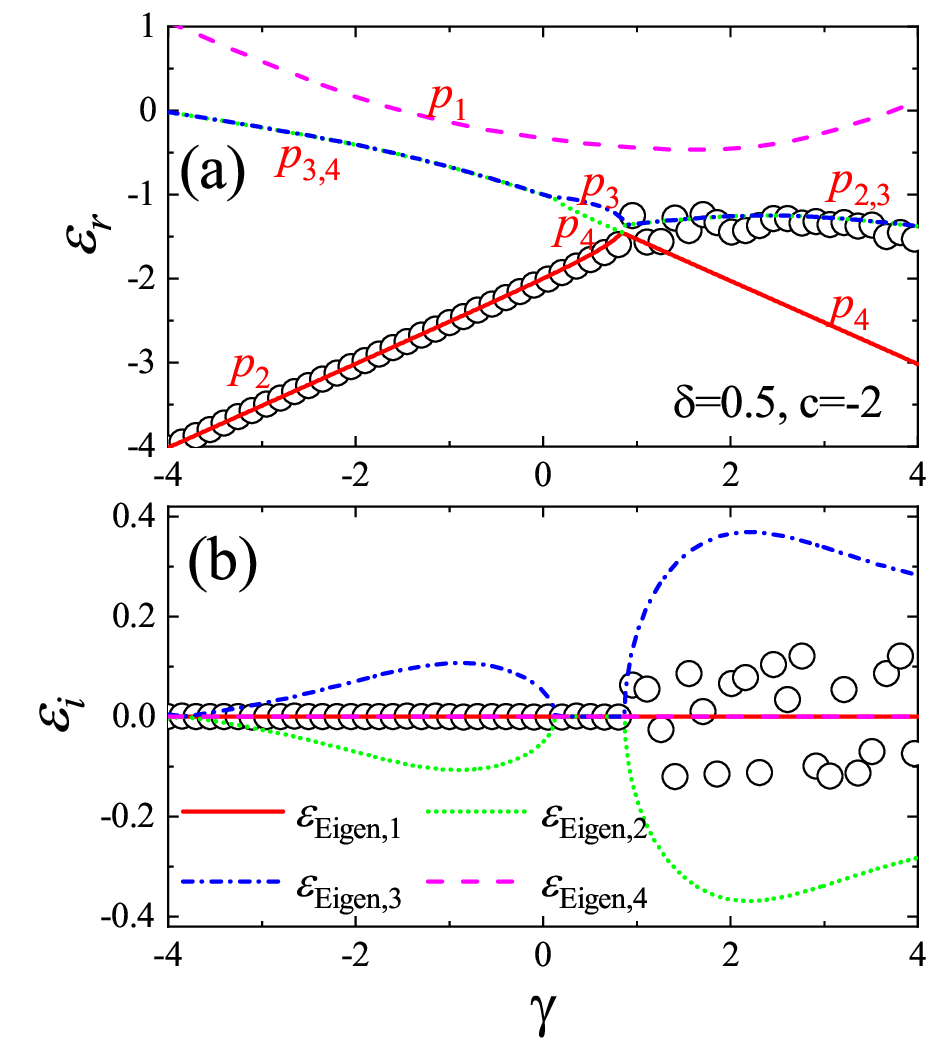}
\caption{(color online) In the presence of nonlinear interactions and in the weak nonreciprocal regime, i.e., $\delta=0.5$, dynamical levels $\varepsilon_{\mathrm{Dyn}}$ (circles) and adiabatic levels $\varepsilon_{\mathrm{Eigen}}$ (solid lines). (a) and (b) present the real part and imaginary part of the energy levels for $c=-2$. For dynamical evolution, the system is initially prepared in the lower adiabatic levels, i.e., $(a,~b)^T=(1,~0)^T$. Here, the dynamic evolution is in the adiabatic limit ($\alpha=0.01$). In (b), $p_i$ are the fixed points shown in Fig. \ref{fig:orbit} corresponding to the eigenstates of the energy level. }
\label{fig:fig7}
\end{figure}
%%%%%%%%%%%%%%%%%%%%%%%%%%%%%%%%%%%%%%%%%%%%%%%%%%%%%%%%%%%%%%%%%%%%%

\section{Adiabaticity of nonreciprocal LZT in the presence of nonlinear interactions ($c\ne 0$)}\label{SecNLinear}
To investigate the adiabaticity, we need to analyze the behavior of the adiabatic levels in the nonlinear model. These levels are the solution of the time-independent version of the Schr\"{o}dinger equation obtained by replacing $i\partial/\partial t$ with the energy $\varepsilon$, i.e., $\varepsilon\left(\begin{array}{l}a \\ b\end{array}\right)=H(\gamma)\left(\begin{array}{l}a \\ b\end{array}\right)$. We assume that initial state is on the lower level, then we obtain a constraint of  $|a|^2+|b|^2/(1-\delta)=1$ according to the invariant $\zeta_+=1$. Then, we find that the eigenenergy $\varepsilon$ satisfies the following quartic equation:
\begin{equation}\label{eq:nonliearlevel}
A\varepsilon^4-B\varepsilon^3+D\varepsilon^2+E\varepsilon+F=0 \;,
\end{equation}
where $A=16(2-\delta)^2$, $B=16(2-\delta)(c (1-\delta)(3-\delta)-\gamma\delta+c)$,\\ $D=8\left(2 c^{2} (1-\delta)((1-\delta)(3 (1-\delta)+7)+3)-\right.$ \\ $\left.\gamma c\delta((1-\delta)(3 (1-\delta)+8)+3)-2 (1-\delta)\left(\gamma^{2}+(1-\delta) v^{2}\right)\right)$, \\
$E=48 c^{3} (1-\delta)^{2}\delta-48 \gamma c^{2} (1-\delta)\left((1-\delta)^{2}-1\right)+$\\$4 c\delta\left(\gamma^{2}((1-\delta)(3 (1-\delta)-8)+3)-4 (1-\delta)^{2} v^{2}\right)+$\\$4 \gamma\left((1-\delta)^{2}-1\right)\left(\gamma^{2}+(1-\delta) v^{2}\right)$, and \\$F=16 c^{4} (1-\delta)^{3}-24 \gamma c^{3}\delta (1-\delta)^{2}-\delta^{2}\left(\gamma^{2}+(1-\delta) v^{2}\right)^{2}+$\\$4 c^{2} (1-\delta)\left(\gamma^{2}((1-\delta)(3 (1-\delta)-7)+3)-4 (1-\delta)^{2} v^{2}\right)-$\\
$2\gamma c\delta\left(\gamma^{2}((3+\delta) (\delta-1)+1)-6 (1-\delta)^{2} v^{2}\right)$.
When $\delta =0$, the above equation reduces to that of \cite{PhysRevA.61.023402}.

\subsection{In the weak nonreciprocal regime ($\delta<1$)}
%%%%%%%%%%%%%%%%%%%%%%%%%%%%%%%%%%%%%%%%%%%%%%%%%%%%%%%%%%%%%%%%%%%%%
\begin{figure}[tb]
\centering
\includegraphics[width=\columnwidth]{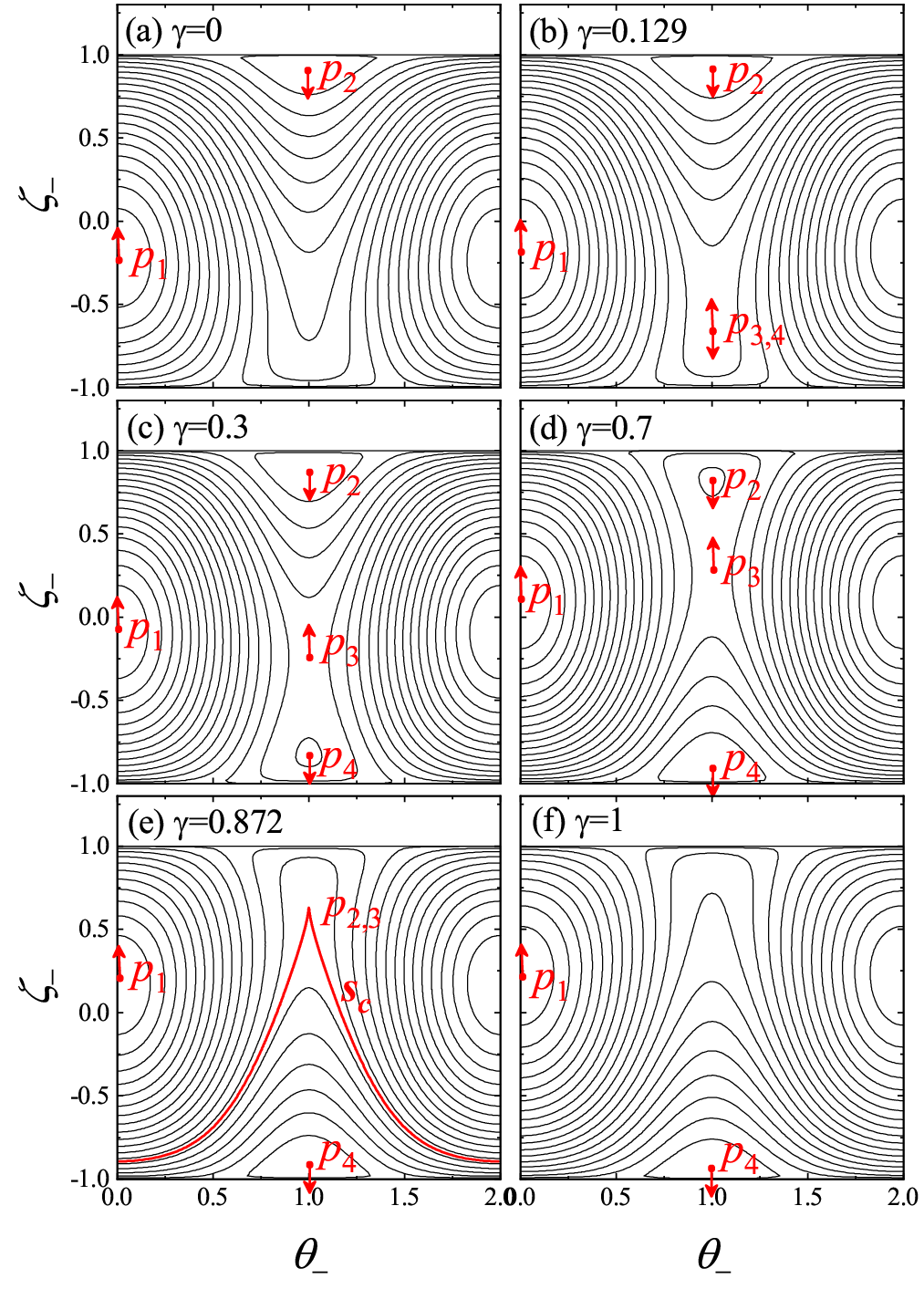}
\caption{(color online) Evolution of the classical trajectories in the plane of two reduced variables $(\theta_-,~\zeta_-)$ for some typical values of $\gamma$. Here, we take $c=-2$ and $\delta=0.5$ as examples. The dots and the bold red line indicate the fixed points. The arrows indicate the shifting direction of the fixed points $p_i$ as $\gamma$ increases. The curves are the periodic trajectories. In (b), we take $\gamma=0.129$, corresponding to the position of $\gamma$ at the right terminal point in Fig. \ref{fig:fig7}(b). In this case, four fixed points occur, implying that four eigenstates do not coalesce. In (e), we take $\gamma=0.872$, corresponding to the left terminal point of the loop in Fig. \ref{fig:fig7}(b). In this case, the fixed points $p_2$ and $p_3$ merge to form a new homoclinic orbit $s_c$, implying that these two eigenstates coalesce. }
\label{fig:orbit}
\end{figure}
%%%%%%%%%%%%%%%%%%%%%%%%%%%%%%%%%%%%%%%%%%%%%%%%%%%%%%%%%%%%%%%%%%%%%

We now focus on the adiabaticity in the weak nonreciprocal regime, i.e., $\delta<1$. We numerically calculate the real and imaginary parts of the energy levels.
In Figs. \ref{fig:fig7}(a) and (b), for a nonlinearity of $c=-2$, we find that the energy levels undergo a dramatic change: there are four pure real energy levels in a window near $\gamma=0$, where a loop appears. Notably, for reciprocal systems, the appearance of a loop breaks the adiabaticity \cite{PhysRevA.61.023402,PhysRevA.61.033603,PhysRevA.66.023404,PhysRevA.66.063603,PhysRevLett.90.170404,Eckel2014}. The primary mechanism for the breakdown of adiabaticity is that when the state moves up to the edge of the loop, it cannot proceed any further except to jump to the upper and lower levels \cite{PhysRevA.61.023402}. However, for the nonreciprocal system that we consider here, as shown in Figs. \ref{fig:fig7}(a) and (b), eigenenergies still exist in the conjugate pair ($p_2$ and $p_3$) beyond the edge of the loop. Nevertheless, the dynamical evolution does not follow these eigenlevels.

To analyze adiabatic dynamics in the presence of nonlinear interactions, in the spirit of Ref. \cite{PhysRevA.66.023404}, we turn to calculate the phase plane of the classical Josephson Hamiltonian \eqref{eq:Hamiltonian1}. In Fig \ref{fig:orbit}, we show the evolution of the classical trajectories in the plane of $(\theta_-,~\zeta_-)$ as $\gamma$ changes adiabatically.
In the plane of $(\theta_-,~\zeta_-)$, the fixed points of the equationsof motion \eqref{eq:dyn1} correspond to the eigenstates that can be obtained by equating the right-hand sides of Eqs. \eqref{eq:dyn1}(b) and (d) to zero.
At the terminal point of the loop, two fixed points merge (as shown in Figs. \ref{fig:orbit}(b) and (e)), indicating that the corresponding eigenstates coalesce. The terminal points of the loop are indeed nonlinearity induced EPs that can also break the adiabaticity of dynamical evolution.

However, at nonlinear EPs, the mechanism of the breakdown of adiabatic evolution is different from that at linear EP. The breakdown of adiabaticity is intimately linked to the collision between the fixed point (i.e., $P_2$ in Fig. 8) and the additional hyperbolic point (i.e., $P_3$ in Fig. 8), where the Bogoliubov excitation frequency drops to zero \cite{PhysRevLett.90.170404}. Moreover,  we find that the invariant $\zeta_+$ does not show a sudden change at the singularity of the nonlinearity-induced EPs.

%%%%%%%%%%%%%%%%%%%%%%%%%%%%%%%%%%%%%%%%%%%%%%%%%%%%%%%%%%%%%%%%%%%%%
\begin{figure}[tb]
\centering
\includegraphics[width=\columnwidth]{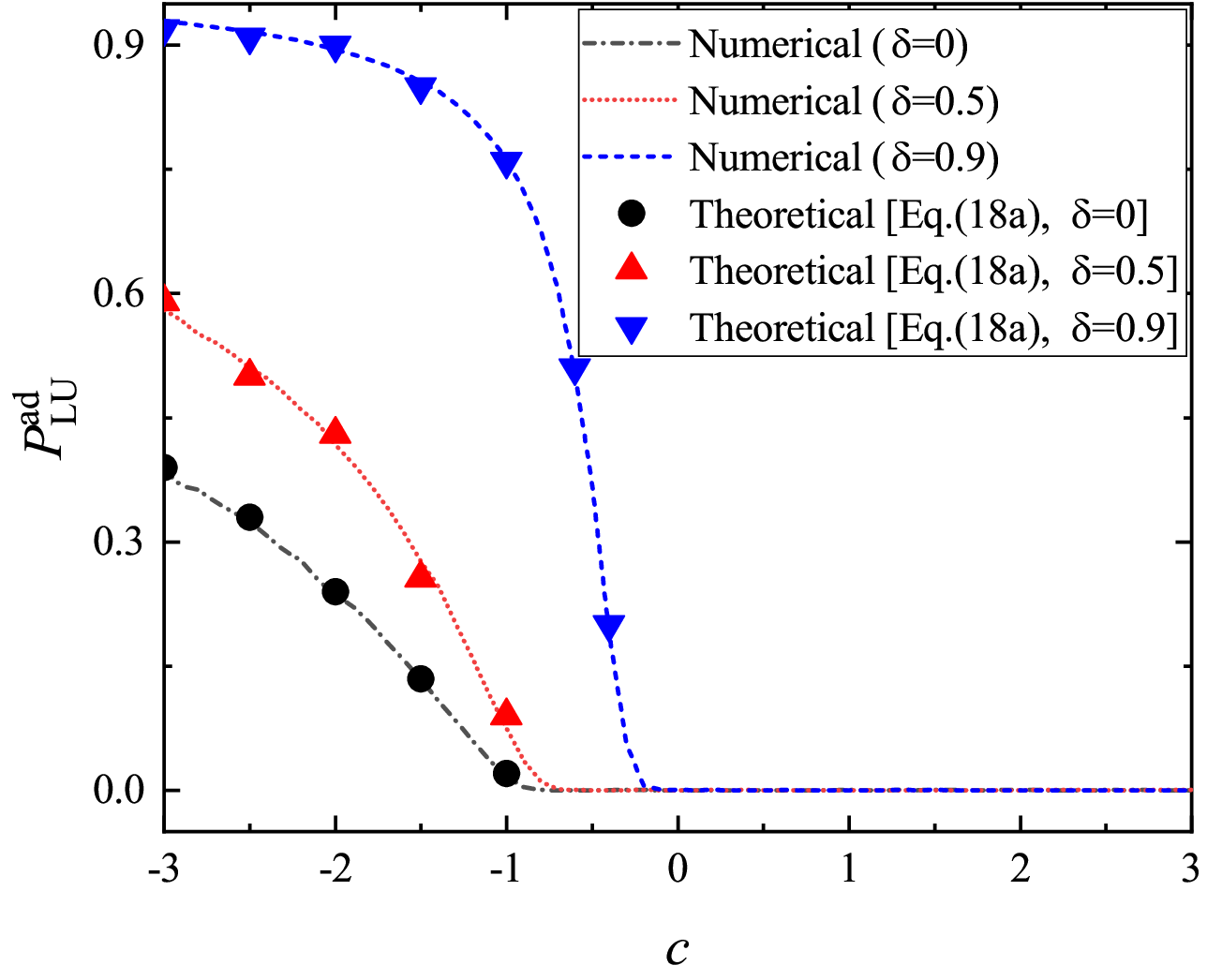}
\caption{(color online) Comparison of the adiabatic tunneling probability $P_{\rm LU}^{\rm ad}$ obtained by directly integrating the time-dependent nonlinear Schr\"{o}dinger equation and the theoretical results obtained using expression \eqref{eq:adiabaticPI}. }
\label{fig:fig9}
\end{figure}
%%%%%%%%%%%%%%%%%%%%%%%%%%%%%%%%%%%%%%%%%%%%%%%%%%%%%%%%%%%%%%%%%%%%%

In Refs. \cite{PhysRevLett.90.170404,PhysRevA.66.023404}, the adiabatic evolution of the fixed points as a function of level bias was proven to correspond to the adiabatic evolution of the eigenstates.
Breakdown of adiabaticity occurs when two fixed points merge together and form a homoclinic orbit.
The nonzero adiabatic tunneling probability can then be calculated from the nonzero classical canonical action ($I_c=\frac{1}{2\pi}\int_0^{2\pi}\zeta_{-}\mathrm{d}\theta_{-}$) of the homoclinic orbit. According to classical adiabatic theorem that action or the area encircled by the classical trajectory is an invariant during the slowly sweeping of $\gamma$. Considering the invariant $\zeta_+=1$ [or $\zeta_+=1/(1-\delta)$] indicating the initial state is on lower (or upper) level, the relation $\zeta_-(t\to+\infty)+1=I_c$ [or $\zeta_-(t\to+\infty)=1/(1-\delta)-I_c$] is obtained. According to Eq. \eqref{eq:pada} [or Eq. \eqref{eq:padb}], we have
\begin{subequations} \label{eq:adiabaticPI}
\begin{align}
&P_{\rm LU}^{\rm ad}=\frac{I_c}{2-2\delta+\delta I_c}, \label{eq:adiabaticPIa}\\
&P_{\rm UL}^{\rm ad}=\frac{(1-\delta)I_c}{\frac{2}{1-\delta}-\delta I_c} .\label{eq:adiabaticPIb}
\end{align}
\end{subequations}

In Fig. \ref{fig:fig9}, we show the adiabatic tunneling probabilities $P_{\rm LU}^{\rm ad}$ obtained by directly integrating the time-dependent nonlinear Schr\"{o}dinger equation and compare them with the above analytical formula calculated from the classical canonical action ($I_c$) of the homoclinic orbit at an EP.
Interestingly we find that they match each other well in the weak nonreciprocal regime ($\delta<1$).

\subsection{In the strong nonreciprocal regime ($\delta>1$)}
The fixed points of the classical Hamiltonian correspond to the eigenstates of the nonlinear two-level system \cite{PhysRevA.66.023404} only  for Hermitian or  weak nonreciprocal systems.
In the strong nonreciprocal regime ($\delta>1$), our numerical calculations show that the eigenstates of the nonlinear nonreciprocal Hamiltonian \eqref{eq:ham} are complex and contain nonzero imaginary parts.
Therefore, in this case, the nonzero adiabatic tunneling probability cannot be viewed as the result of the collision between fixed points of the classical Hamiltonian.
In the strong nonreciprocal regime, the interplay between the  nonreciprocity and nonlinearity will dramatically complicate the  transition dynamics. Therefore, we use a 4-5th-order Runge-Kutta algorithm to trace the quantum evolution numerically and calculate the transition probability for a wide range of nonreciprocal and nonlinear parameters.
%%%%%%%%%%%%%%%%%%%%%%%%%%%%%%%%%%%%%%%%%%%%%%%%%%%%%%%%%%%%%%%%%%%%%
\begin{figure}[tb]
\centering
\includegraphics[width=\columnwidth]{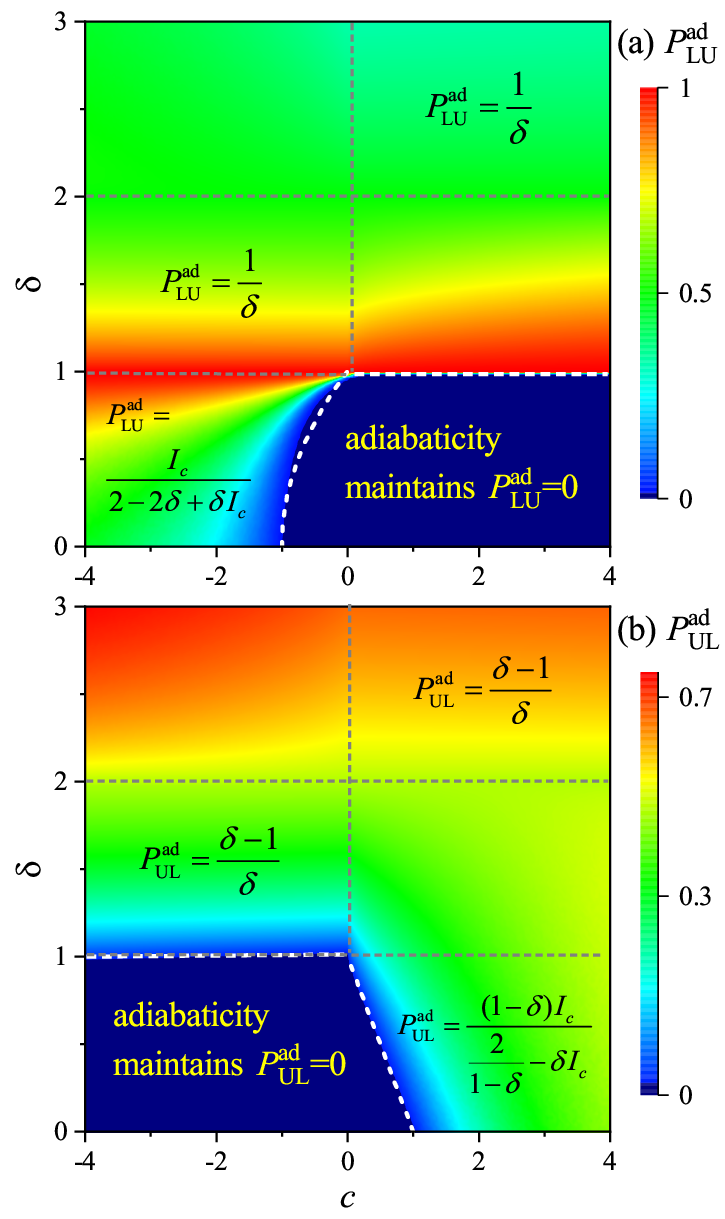}
\caption{(color online) The phase diagram of adiabaticity in the parameter plane ($c,~\delta$). (a) and (b) correspond to adiabatic tunneling probability $P_{\rm LU}^{\rm ad}$ and $P_{\rm UL}^{\rm ad}$, respectively. In the region where adiabaticity is maintained, the adiabatic tunneling probabilities are zero. Here, we still focus on the adiabatic limit and take the sweeping rate $\alpha=0.01$. In (a), the region where adiabaticity is maintained is estimated within parameter range $c>\delta^2-1$ and $\delta<1$. In (b), the region where adiabaticity is maintained is estimated within parameter range $c<1-\delta$ and $\delta\leq1$. In both (a) and (b), in the regions framed by gray dots (estimated within parameter ranges: $1<\delta<2$ for $c\leq0$ and $\delta>2$ for $c\geq0$), the nonzero adiabatic tunneling probabilities are identical to their linear counterparts.  }
\label{fig:fig13}
\end{figure}
%%%%%%%%%%%%%%%%%%%%%%%%%%%%%%%%%%%%%%%%%%%%%%%%%%%%%%%%%%%%%%%%%%%%%

\subsection{Phase diagram of adiabaticity }
Figure \ref{fig:fig13} shows the phase diagram of adiabaticity for a large range of nonreciprocity and nonlinearity parameters, where panels (a) and (b) show the adiabatic transition probabilities $P_{\rm LU}^{\rm ad}$ and $P_{\rm UL}^{\rm ad}$, respectively.
Two regions are observed. In the region where adiabaticity is maintained, the adiabatic tunneling probabilities are zero. Whereas, in the region of the breakdown of adiabaticity, the adiabatic tunneling probabilities are nonzero.
In Fig. \ref{fig:fig13}(a), the region where adiabaticity is maintained is approximately within parameter range $c>\delta^2-1$ and $\delta<1$.
In Fig. \ref{fig:fig13}(b), the region where adiabaticity is maintained is approximately within parameter range $c<1-\delta$ or $\delta\leq1$.
In the weak nonreciprocal regime ($\delta<1$), the nonzero adiabatic tunneling probabilities are explicitly demonstrated by the classical action theory \eqref{eq:adiabaticPI}.
In a certain region of the strong nonreciprocity, i.e.,
$(1<\delta<2; c\leq0)$ and $(\delta>2; c\geq0)$ the regions framed by the gray dots in Fig. \ref{fig:fig13}, we find that interestingly, the nonlinear interaction effects can be completely suppressed so that the nonzero adiabatic tunneling probabilities are identical to their linear counterparts Eq. \eqref{eq:adiabaticP}.
In these regions, our calculations indicate that both total population $N$ and variable $\zeta_{-}$  tend to $\infty$ when $\gamma\to\infty$. Considering the invariant $\zeta_{+}$ and according to Eq. \eqref{eq:pad}, the adiabatic tunneling probabilities actually have the same form as in the linear case.

%%%%%%%%%%%Table begin%%%%%%%%%%%%%%%%%%%%%%%%%%%%%%%%%%%%%%%%%%%
\newcommand{\tabincell}[2]{\begin{tabular}{@{}#1@{}}#2\end{tabular}}
\renewcommand{\arraystretch}{1.5}
\begin{table*}[tb]
  \centering
  \caption{}
  \label{tab:tab1}
\begin{tabular}{c|c|c|c|c|c|c}
  \toprule[0.5pt]
  \toprule[0.5pt]
  \tabincell{c}{Initial state\\ ($\gamma\to-\infty$)}  & Parameter region &  \tabincell{c}{Encounter\\ EPs} & Adiabatic fidelity  & \tabincell{c}{Dynamical energy } & \tabincell{c}{Adiabatic tunneling\\ probability} & Adiabaticity   \\
    \midrule[0.5pt]
  \multirow{2}{*}{lower level} & $\delta<1$ and $c>\delta^2-1$  & no & $-\infty<\gamma<+\infty$, $F^{\rm ad}_{\rm L}=1$ & $-\infty<\gamma<+\infty$, $\varepsilon_{\mathrm{Dyn}}(\alpha\to 0)=\varepsilon_{\mathrm{Eigen}}$ &$P^{\rm ad}_{\rm LU}=0$   & maintain \\ %end line
  \cline{2-7} %short partial horizontal lines from column 2 to column 5
   & $\delta\geq1$ or $c\leq\delta^2-1$  & yes & \tabincell{c}{$\gamma\leq+\gamma_{\rm EP}$, $F^{\rm ad}_{\rm L}=1$;\\ $\gamma>+\gamma_{\rm EP}$, $F^{\rm ad}_{\rm L}\neq1$ }  &  \tabincell{c}{$\gamma\leq+\gamma_{\rm EP}$, $\varepsilon_{\mathrm{Dyn}}(\alpha\to 0)=\varepsilon_{\mathrm{Eigen}}$; \\ $\gamma>+\gamma_{\rm EP}$, $\varepsilon_{\mathrm{Dyn}}(\alpha\to 0)\neq\varepsilon_{\mathrm{Eigen}}$  }  & $P^{\rm ad}_{\rm LU}\neq0$  & breakdown \\ %first cell is occupied by the multirow
  \midrule[0.5pt]
  \multirow{2}{*}{upper level} & $\delta\leq1$ and $c<1-\delta$ & no & $-\infty<\gamma<+\infty$, $F^{\rm ad}_{\rm U}=1$ &  $-\infty<\gamma<+\infty$, $\varepsilon_{\mathrm{Dyn}}(\alpha\to 0)=\varepsilon_{\mathrm{Eigen}}$ &  $P^{\rm ad}_{\rm UL}=0$  & maintain \\ %end line
  \cline{2-7} %short partial horizontal lines from column 2 to column 5
   & $\delta>1$ or $c\leq1-\delta$ & yes & \tabincell{c}{$\gamma\leq-\gamma_{\rm EP}$, $F^{\rm ad}_{\rm U}=1$;\\ $\gamma>-\gamma_{\rm EP}$, $F^{\rm ad}_{\rm U}\neq1$ } &  \tabincell{c}{$\gamma\leq-\gamma_{\rm EP}$, $\varepsilon_{\mathrm{Dyn}}(\alpha\to 0)=\varepsilon_{\mathrm{Eigen}}$; \\ $\gamma>-\gamma_{\rm EP}$, $\varepsilon_{\mathrm{Dyn}}(\alpha\to 0)\neq\varepsilon_{\mathrm{Eigen}}$  } & $P^{\rm ad}_{\rm UL}\neq0$ & breakdown \\ %first cell is occupied by the multirow
  \bottomrule[0.5pt]
  \bottomrule[0.5pt]
\end{tabular}
\end{table*}
%%%%%%%%%%%Table end%%%%%%%%%%%%%%%%%%%%%%%%%%%%%%%%%%%%%%%%%%%

\section{Some remarks}\label{SecDiscussions}
For the non-Hermitian systems, the distinct feature is the appearance of complex eigenvalues. Thus, the time evolution of such a non-Hermitian system is no longer unitary.
To trace the temporal evolution of the probability distribution, a proper normalization scheme is required. Two approaches are usually used for the normalization of a quantum state:
Scheme I: normalization of  the amplitudes of the wavefunction only at the end of time evolution \cite{PhysRevA.102.062213}; Scheme II: normalization of the amplitudes of the wavefunction in each integration step \cite{PhysRevA.98.022129,PhysRevB.98.205417,PhysRevB.103.054203}.

Taking the  present work on nonreciprocal LZT for example,
in the linear case of $c=0$, we find that the results obtained from the two normalization schemes are completely consistent with each other for both the adiabatic case and the nonadiabatic case.  While for the nonlinear case of $c\neq0$,
the two normalization schemes usually lead to different results.
Moreover, the results from Scheme II are usually dependent on the choice of integration step. For the adiabatic system, Scheme II can give a convergent adiabatic tunneling probability when the  integration step is set to be small enough.

For most  current experiments \cite{El-Ganainy2018,doi:10.1126/science.aar7709,doi:10.1080/00018732.2021.1876991,Bouganne2020,RevModPhys.93.015005},
 the normalization of the wavefunction is not performed  at each time step, but rather is performed at the end of the time evolution. That is, most experiments are mainly concerned with the  fractions of the populations on different levels at the end of a non-unitary evolution of a non-Hermitian quantum system. Therefore, in the present work,
we choose to normalize the amplitudes of the wavefunction only at the end of time evolution.

\section{Conclusion }\label{SecCon}
In this work, we numerically and analytically investigate adiabaticity in the nonreciprocal Landau-Zener model with a nonlinear self-interaction.
In this model, hopping nonreciprocity can lead to non-Hermiticity, which is distinct from the $\mathcal{PT}$ system for which non-Hermiticity is induced by a complex onsite potential.
For the first time, we derive the Lagrangian and classical Josephson Hamiltonian of the nonreciprocal non-Hermitian system.
In the adiabatic limit of parameter sweeping, a quantum state is found to still follow the eigenstate solution until it encounters the EPs that dramatically alter the tunneling process, leading to the breakdown of adiabaticity.
The EPs can be formed and changed by both nonreciprocity and nonlinearity parameters. The competition between nonreciprocity and nonlinearity yields a rich variety of adiabatic quantum evolutions.

An explicit phase diagram of adiabaticity for a large range of nonreciprocity and nonlinearity parameters is obtained. The main results are summarized in Table \ref{tab:tab1}. Our work provides a non-Hermitian extension of the celebrated LZT model, achieves insight into the adiabatic evolution of the non-Hermitian system, and may stimulate further explorations of the non-Hermitian topological properties both theoretically and experimentally.

\section*{Acknowledgments}
We are grateful to Difa Ye for valuable suggestions and discussions. This work was supported by the National Natural Science Foundation of China (Contract No. 12005173), the Natural Science Foundation of Gansu Province (Contract No. 20JR10RA082), the China Postdoctoral Science Foundation (Contract No. 2020M680318), and the NSAF (Contract No. U1930402, No. U1930403).

%\bibliographystyle{plain}
%\bibliography{NLZ}

\begin{thebibliography}{80}%
\makeatletter
\providecommand \@ifxundefined [1]{%
 \@ifx{#1\undefined}
}%
\providecommand \@ifnum [1]{%
 \ifnum #1\expandafter \@firstoftwo
 \else \expandafter \@secondoftwo
 \fi
}%
\providecommand \@ifx [1]{%
 \ifx #1\expandafter \@firstoftwo
 \else \expandafter \@secondoftwo
 \fi
}%
\providecommand \natexlab [1]{#1}%
\providecommand \enquote  [1]{``#1''}%
\providecommand \bibnamefont  [1]{#1}%
\providecommand \bibfnamefont [1]{#1}%
\providecommand \citenamefont [1]{#1}%
\providecommand \href@noop [0]{\@secondoftwo}%
\providecommand \href [0]{\begingroup \@sanitize@url \@href}%
\providecommand \@href[1]{\@@startlink{#1}\@@href}%
\providecommand \@@href[1]{\endgroup#1\@@endlink}%
\providecommand \@sanitize@url [0]{\catcode `\\12\catcode `\$12\catcode
  `\&12\catcode `\#12\catcode `\^12\catcode `\_12\catcode `\%12\relax}%
\providecommand \@@startlink[1]{}%
\providecommand \@@endlink[0]{}%
\providecommand \url  [0]{\begingroup\@sanitize@url \@url }%
\providecommand \@url [1]{\endgroup\@href {#1}{\urlprefix }}%
\providecommand \urlprefix  [0]{URL }%
\providecommand \Eprint [0]{\href }%
\providecommand \doibase [0]{http://dx.doi.org/}%
\providecommand \selectlanguage [0]{\@gobble}%
\providecommand \bibinfo  [0]{\@secondoftwo}%
\providecommand \bibfield  [0]{\@secondoftwo}%
\providecommand \translation [1]{[#1]}%
\providecommand \BibitemOpen [0]{}%
\providecommand \bibitemStop [0]{}%
\providecommand \bibitemNoStop [0]{.\EOS\space}%
\providecommand \EOS [0]{\spacefactor3000\relax}%
\providecommand \BibitemShut  [1]{\csname bibitem#1\endcsname}%
\let\auto@bib@innerbib\@empty
%</preamble>
\bibitem [{\citenamefont {Landau}(1932)}]{Landau1932}%
  \BibitemOpen
  \bibfield  {author} {\bibinfo {author} {\bibfnamefont {L.~D.}\ \bibnamefont
  {Landau}},\ }\href@noop {} {\bibfield  {journal} {\bibinfo  {journal} {Phys.
  Z. Sowjetunion}\ }\textbf {\bibinfo {volume} {2}},\ \bibinfo {pages} {46}
  (\bibinfo {year} {1932})}\BibitemShut {NoStop}%
\bibitem [{\citenamefont {Zener}(1932)}]{Zener1932}%
  \BibitemOpen
  \bibfield  {author} {\bibinfo {author} {\bibfnamefont {C.}~\bibnamefont
  {Zener}},\ }\href {\doibase 10.1098/rspa.1932.0165} {\bibfield  {journal}
  {\bibinfo  {journal} {Proc. R. Soc. London, Ser. A}\ }\textbf {\bibinfo
  {volume} {137}},\ \bibinfo {pages} {696} (\bibinfo {year}
  {1932})}\BibitemShut {NoStop}%
\bibitem [{\citenamefont {Sillanp\"a\"a}\ \emph {et~al.}(2006)\citenamefont
  {Sillanp\"a\"a}, \citenamefont {Lehtinen}, \citenamefont {Paila},
  \citenamefont {Makhlin},\ and\ \citenamefont
  {Hakonen}}]{PhysRevLett.96.187002}%
  \BibitemOpen
  \bibfield  {author} {\bibinfo {author} {\bibfnamefont {M.}~\bibnamefont
  {Sillanp\"a\"a}}, \bibinfo {author} {\bibfnamefont {T.}~\bibnamefont
  {Lehtinen}}, \bibinfo {author} {\bibfnamefont {A.}~\bibnamefont {Paila}},
  \bibinfo {author} {\bibfnamefont {Y.}~\bibnamefont {Makhlin}}, \ and\
  \bibinfo {author} {\bibfnamefont {P.}~\bibnamefont {Hakonen}},\ }\href
  {\doibase 10.1103/PhysRevLett.96.187002} {\bibfield  {journal} {\bibinfo
  {journal} {Phys. Rev. Lett.}\ }\textbf {\bibinfo {volume} {96}},\ \bibinfo
  {pages} {187002} (\bibinfo {year} {2006})}\BibitemShut {NoStop}%
\bibitem [{\citenamefont {Ashhab}\ \emph {et~al.}(2006)\citenamefont {Ashhab},
  \citenamefont {Johansson},\ and\ \citenamefont {Nori}}]{PhysRevA.74.052330}%
  \BibitemOpen
  \bibfield  {author} {\bibinfo {author} {\bibfnamefont {S.}~\bibnamefont
  {Ashhab}}, \bibinfo {author} {\bibfnamefont {J.~R.}\ \bibnamefont
  {Johansson}}, \ and\ \bibinfo {author} {\bibfnamefont {F.}~\bibnamefont
  {Nori}},\ }\href {\doibase 10.1103/PhysRevA.74.052330} {\bibfield  {journal}
  {\bibinfo  {journal} {Phys. Rev. A}\ }\textbf {\bibinfo {volume} {74}},\
  \bibinfo {pages} {052330} (\bibinfo {year} {2006})}\BibitemShut {NoStop}%
\bibitem [{\citenamefont {Berns}\ \emph {et~al.}(2008)\citenamefont {Berns},
  \citenamefont {Rudner}, \citenamefont {Valenzuela}, \citenamefont {Berggren},
  \citenamefont {Oliver}, \citenamefont {Levitov},\ and\ \citenamefont
  {Orlando}}]{Berns2008}%
  \BibitemOpen
  \bibfield  {author} {\bibinfo {author} {\bibfnamefont {D.~M.}\ \bibnamefont
  {Berns}}, \bibinfo {author} {\bibfnamefont {M.~S.}\ \bibnamefont {Rudner}},
  \bibinfo {author} {\bibfnamefont {S.~O.}\ \bibnamefont {Valenzuela}},
  \bibinfo {author} {\bibfnamefont {K.~K.}\ \bibnamefont {Berggren}}, \bibinfo
  {author} {\bibfnamefont {W.~D.}\ \bibnamefont {Oliver}}, \bibinfo {author}
  {\bibfnamefont {L.~S.}\ \bibnamefont {Levitov}}, \ and\ \bibinfo {author}
  {\bibfnamefont {T.~P.}\ \bibnamefont {Orlando}},\ }\href {\doibase
  10.1038/nature07262} {\bibfield  {journal} {\bibinfo  {journal} {Nature}\
  }\textbf {\bibinfo {volume} {455}},\ \bibinfo {pages} {51} (\bibinfo {year}
  {2008})}\BibitemShut {NoStop}%
\bibitem [{\citenamefont {Fuchs}\ \emph {et~al.}(2009)\citenamefont {Fuchs},
  \citenamefont {Dobrovitski}, \citenamefont {Toyli}, \citenamefont
  {Heremans},\ and\ \citenamefont {Awschalom}}]{doi:10.1126/science.1181193}%
  \BibitemOpen
  \bibfield  {author} {\bibinfo {author} {\bibfnamefont {G.~D.}\ \bibnamefont
  {Fuchs}}, \bibinfo {author} {\bibfnamefont {V.~V.}\ \bibnamefont
  {Dobrovitski}}, \bibinfo {author} {\bibfnamefont {D.~M.}\ \bibnamefont
  {Toyli}}, \bibinfo {author} {\bibfnamefont {F.~J.}\ \bibnamefont {Heremans}},
  \ and\ \bibinfo {author} {\bibfnamefont {D.~D.}\ \bibnamefont {Awschalom}},\
  }\href {\doibase 10.1126/science.1181193} {\bibfield  {journal} {\bibinfo
  {journal} {Science}\ }\textbf {\bibinfo {volume} {326}},\ \bibinfo {pages}
  {1520} (\bibinfo {year} {2009})}\BibitemShut {NoStop}%
\bibitem [{\citenamefont {Ribeiro}\ and\ \citenamefont
  {Burkard}(2009)}]{PhysRevLett.102.216802}%
  \BibitemOpen
  \bibfield  {author} {\bibinfo {author} {\bibfnamefont {H.}~\bibnamefont
  {Ribeiro}}\ and\ \bibinfo {author} {\bibfnamefont {G.}~\bibnamefont
  {Burkard}},\ }\href {\doibase 10.1103/PhysRevLett.102.216802} {\bibfield
  {journal} {\bibinfo  {journal} {Phys. Rev. Lett.}\ }\textbf {\bibinfo
  {volume} {102}},\ \bibinfo {pages} {216802} (\bibinfo {year}
  {2009})}\BibitemShut {NoStop}%
\bibitem [{\citenamefont {Petta}\ \emph {et~al.}(2010)\citenamefont {Petta},
  \citenamefont {Lu},\ and\ \citenamefont
  {Gossard}}]{doi:10.1126/science.1183628}%
  \BibitemOpen
  \bibfield  {author} {\bibinfo {author} {\bibfnamefont {J.~R.}\ \bibnamefont
  {Petta}}, \bibinfo {author} {\bibfnamefont {H.}~\bibnamefont {Lu}}, \ and\
  \bibinfo {author} {\bibfnamefont {A.~C.}\ \bibnamefont {Gossard}},\ }\href
  {\doibase 10.1126/science.1183628} {\bibfield  {journal} {\bibinfo  {journal}
  {Science}\ }\textbf {\bibinfo {volume} {327}},\ \bibinfo {pages} {669}
  (\bibinfo {year} {2010})}\BibitemShut {NoStop}%
\bibitem [{\citenamefont {Khomeriki}\ and\ \citenamefont
  {Ruffo}(2005)}]{PhysRevLett.94.113904}%
  \BibitemOpen
  \bibfield  {author} {\bibinfo {author} {\bibfnamefont {R.}~\bibnamefont
  {Khomeriki}}\ and\ \bibinfo {author} {\bibfnamefont {S.}~\bibnamefont
  {Ruffo}},\ }\href {\doibase 10.1103/PhysRevLett.94.113904} {\bibfield
  {journal} {\bibinfo  {journal} {Phys. Rev. Lett.}\ }\textbf {\bibinfo
  {volume} {94}},\ \bibinfo {pages} {113904} (\bibinfo {year}
  {2005})}\BibitemShut {NoStop}%
\bibitem [{\citenamefont {Wu}\ and\ \citenamefont
  {Niu}(2000)}]{PhysRevA.61.023402}%
  \BibitemOpen
  \bibfield  {author} {\bibinfo {author} {\bibfnamefont {B.}~\bibnamefont
  {Wu}}\ and\ \bibinfo {author} {\bibfnamefont {Q.}~\bibnamefont {Niu}},\
  }\href {\doibase 10.1103/PhysRevA.61.023402} {\bibfield  {journal} {\bibinfo
  {journal} {Phys. Rev. A}\ }\textbf {\bibinfo {volume} {61}},\ \bibinfo
  {pages} {023402} (\bibinfo {year} {2000})}\BibitemShut {NoStop}%
\bibitem [{\citenamefont {Zobay}\ and\ \citenamefont
  {Garraway}(2000)}]{PhysRevA.61.033603}%
  \BibitemOpen
  \bibfield  {author} {\bibinfo {author} {\bibfnamefont {O.}~\bibnamefont
  {Zobay}}\ and\ \bibinfo {author} {\bibfnamefont {B.~M.}\ \bibnamefont
  {Garraway}},\ }\href {\doibase 10.1103/PhysRevA.61.033603} {\bibfield
  {journal} {\bibinfo  {journal} {Phys. Rev. A}\ }\textbf {\bibinfo {volume}
  {61}},\ \bibinfo {pages} {033603} (\bibinfo {year} {2000})}\BibitemShut
  {NoStop}%
\bibitem [{\citenamefont {Liu}\ \emph {et~al.}(2002)\citenamefont {Liu},
  \citenamefont {Fu}, \citenamefont {Ou}, \citenamefont {Chen}, \citenamefont
  {Choi}, \citenamefont {Wu},\ and\ \citenamefont {Niu}}]{PhysRevA.66.023404}%
  \BibitemOpen
  \bibfield  {author} {\bibinfo {author} {\bibfnamefont {J.}~\bibnamefont
  {Liu}}, \bibinfo {author} {\bibfnamefont {L.}~\bibnamefont {Fu}}, \bibinfo
  {author} {\bibfnamefont {B.-Y.}\ \bibnamefont {Ou}}, \bibinfo {author}
  {\bibfnamefont {S.-G.}\ \bibnamefont {Chen}}, \bibinfo {author}
  {\bibfnamefont {D.-I.}\ \bibnamefont {Choi}}, \bibinfo {author}
  {\bibfnamefont {B.}~\bibnamefont {Wu}}, \ and\ \bibinfo {author}
  {\bibfnamefont {Q.}~\bibnamefont {Niu}},\ }\href {\doibase
  10.1103/PhysRevA.66.023404} {\bibfield  {journal} {\bibinfo  {journal} {Phys.
  Rev. A}\ }\textbf {\bibinfo {volume} {66}},\ \bibinfo {pages} {023404}
  (\bibinfo {year} {2002})}\BibitemShut {NoStop}%
\bibitem [{\citenamefont {Mueller}(2002)}]{PhysRevA.66.063603}%
  \BibitemOpen
  \bibfield  {author} {\bibinfo {author} {\bibfnamefont {E.~J.}\ \bibnamefont
  {Mueller}},\ }\href {\doibase 10.1103/PhysRevA.66.063603} {\bibfield
  {journal} {\bibinfo  {journal} {Phys. Rev. A}\ }\textbf {\bibinfo {volume}
  {66}},\ \bibinfo {pages} {063603} (\bibinfo {year} {2002})}\BibitemShut
  {NoStop}%
\bibitem [{\citenamefont {Morsch}\ \emph {et~al.}(2001)\citenamefont {Morsch},
  \citenamefont {M\"uller}, \citenamefont {Cristiani}, \citenamefont
  {Ciampini},\ and\ \citenamefont {Arimondo}}]{PhysRevLett.87.140402}%
  \BibitemOpen
  \bibfield  {author} {\bibinfo {author} {\bibfnamefont {O.}~\bibnamefont
  {Morsch}}, \bibinfo {author} {\bibfnamefont {J.~H.}\ \bibnamefont
  {M\"uller}}, \bibinfo {author} {\bibfnamefont {M.}~\bibnamefont {Cristiani}},
  \bibinfo {author} {\bibfnamefont {D.}~\bibnamefont {Ciampini}}, \ and\
  \bibinfo {author} {\bibfnamefont {E.}~\bibnamefont {Arimondo}},\ }\href
  {\doibase 10.1103/PhysRevLett.87.140402} {\bibfield  {journal} {\bibinfo
  {journal} {Phys. Rev. Lett.}\ }\textbf {\bibinfo {volume} {87}},\ \bibinfo
  {pages} {140402} (\bibinfo {year} {2001})}\BibitemShut {NoStop}%
\bibitem [{\citenamefont {Cristiani}\ \emph {et~al.}(2002)\citenamefont
  {Cristiani}, \citenamefont {Morsch}, \citenamefont {M\"uller}, \citenamefont
  {Ciampini},\ and\ \citenamefont {Arimondo}}]{PhysRevA.65.063612}%
  \BibitemOpen
  \bibfield  {author} {\bibinfo {author} {\bibfnamefont {M.}~\bibnamefont
  {Cristiani}}, \bibinfo {author} {\bibfnamefont {O.}~\bibnamefont {Morsch}},
  \bibinfo {author} {\bibfnamefont {J.~H.}\ \bibnamefont {M\"uller}}, \bibinfo
  {author} {\bibfnamefont {D.}~\bibnamefont {Ciampini}}, \ and\ \bibinfo
  {author} {\bibfnamefont {E.}~\bibnamefont {Arimondo}},\ }\href {\doibase
  10.1103/PhysRevA.65.063612} {\bibfield  {journal} {\bibinfo  {journal} {Phys.
  Rev. A}\ }\textbf {\bibinfo {volume} {65}},\ \bibinfo {pages} {063612}
  (\bibinfo {year} {2002})}\BibitemShut {NoStop}%
\bibitem [{\citenamefont {Liu}\ \emph {et~al.}(2003)\citenamefont {Liu},
  \citenamefont {Wu},\ and\ \citenamefont {Niu}}]{PhysRevLett.90.170404}%
  \BibitemOpen
  \bibfield  {author} {\bibinfo {author} {\bibfnamefont {J.}~\bibnamefont
  {Liu}}, \bibinfo {author} {\bibfnamefont {B.}~\bibnamefont {Wu}}, \ and\
  \bibinfo {author} {\bibfnamefont {Q.}~\bibnamefont {Niu}},\ }\href {\doibase
  10.1103/PhysRevLett.90.170404} {\bibfield  {journal} {\bibinfo  {journal}
  {Phys. Rev. Lett.}\ }\textbf {\bibinfo {volume} {90}},\ \bibinfo {pages}
  {170404} (\bibinfo {year} {2003})}\BibitemShut {NoStop}%
\bibitem [{\citenamefont {Jona-Lasinio}\ \emph {et~al.}(2003)\citenamefont
  {Jona-Lasinio}, \citenamefont {Morsch}, \citenamefont {Cristiani},
  \citenamefont {Malossi}, \citenamefont {M\"uller}, \citenamefont {Courtade},
  \citenamefont {Anderlini},\ and\ \citenamefont
  {Arimondo}}]{PhysRevLett.91.230406}%
  \BibitemOpen
  \bibfield  {author} {\bibinfo {author} {\bibfnamefont {M.}~\bibnamefont
  {Jona-Lasinio}}, \bibinfo {author} {\bibfnamefont {O.}~\bibnamefont
  {Morsch}}, \bibinfo {author} {\bibfnamefont {M.}~\bibnamefont {Cristiani}},
  \bibinfo {author} {\bibfnamefont {N.}~\bibnamefont {Malossi}}, \bibinfo
  {author} {\bibfnamefont {J.~H.}\ \bibnamefont {M\"uller}}, \bibinfo {author}
  {\bibfnamefont {E.}~\bibnamefont {Courtade}}, \bibinfo {author}
  {\bibfnamefont {M.}~\bibnamefont {Anderlini}}, \ and\ \bibinfo {author}
  {\bibfnamefont {E.}~\bibnamefont {Arimondo}},\ }\href {\doibase
  10.1103/PhysRevLett.91.230406} {\bibfield  {journal} {\bibinfo  {journal}
  {Phys. Rev. Lett.}\ }\textbf {\bibinfo {volume} {91}},\ \bibinfo {pages}
  {230406} (\bibinfo {year} {2003})}\BibitemShut {NoStop}%
\bibitem [{\citenamefont {Witthaut}\ \emph {et~al.}(2006)\citenamefont
  {Witthaut}, \citenamefont {Graefe},\ and\ \citenamefont
  {Korsch}}]{PhysRevA.73.063609}%
  \BibitemOpen
  \bibfield  {author} {\bibinfo {author} {\bibfnamefont {D.}~\bibnamefont
  {Witthaut}}, \bibinfo {author} {\bibfnamefont {E.~M.}\ \bibnamefont
  {Graefe}}, \ and\ \bibinfo {author} {\bibfnamefont {H.~J.}\ \bibnamefont
  {Korsch}},\ }\href {\doibase 10.1103/PhysRevA.73.063609} {\bibfield
  {journal} {\bibinfo  {journal} {Phys. Rev. A}\ }\textbf {\bibinfo {volume}
  {73}},\ \bibinfo {pages} {063609} (\bibinfo {year} {2006})}\BibitemShut
  {NoStop}%
\bibitem [{\citenamefont {Wu}\ and\ \citenamefont
  {Liu}(2006)}]{PhysRevLett.96.020405}%
  \BibitemOpen
  \bibfield  {author} {\bibinfo {author} {\bibfnamefont {B.}~\bibnamefont
  {Wu}}\ and\ \bibinfo {author} {\bibfnamefont {J.}~\bibnamefont {Liu}},\
  }\href {\doibase 10.1103/PhysRevLett.96.020405} {\bibfield  {journal}
  {\bibinfo  {journal} {Phys. Rev. Lett.}\ }\textbf {\bibinfo {volume} {96}},\
  \bibinfo {pages} {020405} (\bibinfo {year} {2006})}\BibitemShut {NoStop}%
\bibitem [{\citenamefont {Smith-Mannschott}\ \emph {et~al.}(2009)\citenamefont
  {Smith-Mannschott}, \citenamefont {Chuchem}, \citenamefont {Hiller},
  \citenamefont {Kottos},\ and\ \citenamefont
  {Cohen}}]{PhysRevLett.102.230401}%
  \BibitemOpen
  \bibfield  {author} {\bibinfo {author} {\bibfnamefont {K.}~\bibnamefont
  {Smith-Mannschott}}, \bibinfo {author} {\bibfnamefont {M.}~\bibnamefont
  {Chuchem}}, \bibinfo {author} {\bibfnamefont {M.}~\bibnamefont {Hiller}},
  \bibinfo {author} {\bibfnamefont {T.}~\bibnamefont {Kottos}}, \ and\ \bibinfo
  {author} {\bibfnamefont {D.}~\bibnamefont {Cohen}},\ }\href {\doibase
  10.1103/PhysRevLett.102.230401} {\bibfield  {journal} {\bibinfo  {journal}
  {Phys. Rev. Lett.}\ }\textbf {\bibinfo {volume} {102}},\ \bibinfo {pages}
  {230401} (\bibinfo {year} {2009})}\BibitemShut {NoStop}%
\bibitem [{\citenamefont {Zenesini}\ \emph {et~al.}(2008)\citenamefont
  {Zenesini}, \citenamefont {Sias}, \citenamefont {Lignier}, \citenamefont
  {Singh}, \citenamefont {Ciampini}, \citenamefont {Morsch}, \citenamefont
  {Mannella}, \citenamefont {Arimondo}, \citenamefont {Tomadin},\ and\
  \citenamefont {Wimberger}}]{Zenesini_2008}%
  \BibitemOpen
  \bibfield  {author} {\bibinfo {author} {\bibfnamefont {A.}~\bibnamefont
  {Zenesini}}, \bibinfo {author} {\bibfnamefont {C.}~\bibnamefont {Sias}},
  \bibinfo {author} {\bibfnamefont {H.}~\bibnamefont {Lignier}}, \bibinfo
  {author} {\bibfnamefont {Y.}~\bibnamefont {Singh}}, \bibinfo {author}
  {\bibfnamefont {D.}~\bibnamefont {Ciampini}}, \bibinfo {author}
  {\bibfnamefont {O.}~\bibnamefont {Morsch}}, \bibinfo {author} {\bibfnamefont
  {R.}~\bibnamefont {Mannella}}, \bibinfo {author} {\bibfnamefont
  {E.}~\bibnamefont {Arimondo}}, \bibinfo {author} {\bibfnamefont
  {A.}~\bibnamefont {Tomadin}}, \ and\ \bibinfo {author} {\bibfnamefont
  {S.}~\bibnamefont {Wimberger}},\ }\href {\doibase
  10.1088/1367-2630/10/5/053038} {\bibfield  {journal} {\bibinfo  {journal}
  {New J. Phys.}\ }\textbf {\bibinfo {volume} {10}},\ \bibinfo {pages} {053038}
  (\bibinfo {year} {2008})}\BibitemShut {NoStop}%
\bibitem [{\citenamefont {Zenesini}\ \emph {et~al.}(2009)\citenamefont
  {Zenesini}, \citenamefont {Lignier}, \citenamefont {Tayebirad}, \citenamefont
  {Radogostowicz}, \citenamefont {Ciampini}, \citenamefont {Mannella},
  \citenamefont {Wimberger}, \citenamefont {Morsch},\ and\ \citenamefont
  {Arimondo}}]{PhysRevLett.103.090403}%
  \BibitemOpen
  \bibfield  {author} {\bibinfo {author} {\bibfnamefont {A.}~\bibnamefont
  {Zenesini}}, \bibinfo {author} {\bibfnamefont {H.}~\bibnamefont {Lignier}},
  \bibinfo {author} {\bibfnamefont {G.}~\bibnamefont {Tayebirad}}, \bibinfo
  {author} {\bibfnamefont {J.}~\bibnamefont {Radogostowicz}}, \bibinfo {author}
  {\bibfnamefont {D.}~\bibnamefont {Ciampini}}, \bibinfo {author}
  {\bibfnamefont {R.}~\bibnamefont {Mannella}}, \bibinfo {author}
  {\bibfnamefont {S.}~\bibnamefont {Wimberger}}, \bibinfo {author}
  {\bibfnamefont {O.}~\bibnamefont {Morsch}}, \ and\ \bibinfo {author}
  {\bibfnamefont {E.}~\bibnamefont {Arimondo}},\ }\href {\doibase
  10.1103/PhysRevLett.103.090403} {\bibfield  {journal} {\bibinfo  {journal}
  {Phys. Rev. Lett.}\ }\textbf {\bibinfo {volume} {103}},\ \bibinfo {pages}
  {090403} (\bibinfo {year} {2009})}\BibitemShut {NoStop}%
\bibitem [{\citenamefont {Chen}\ \emph {et~al.}(2011)\citenamefont {Chen},
  \citenamefont {Huber}, \citenamefont {Trotzky}, \citenamefont {Bloch},\ and\
  \citenamefont {Altman}}]{Chen2011}%
  \BibitemOpen
  \bibfield  {author} {\bibinfo {author} {\bibfnamefont {Y.-A.}\ \bibnamefont
  {Chen}}, \bibinfo {author} {\bibfnamefont {S.~D.}\ \bibnamefont {Huber}},
  \bibinfo {author} {\bibfnamefont {S.}~\bibnamefont {Trotzky}}, \bibinfo
  {author} {\bibfnamefont {I.}~\bibnamefont {Bloch}}, \ and\ \bibinfo {author}
  {\bibfnamefont {E.}~\bibnamefont {Altman}},\ }\href {\doibase
  10.1038/nphys1801} {\bibfield  {journal} {\bibinfo  {journal} {Phys. Rev.
  Lett.}\ }\textbf {\bibinfo {volume} {7}},\ \bibinfo {pages} {61} (\bibinfo
  {year} {2011})}\BibitemShut {NoStop}%
\bibitem [{\citenamefont {Kasztelan}\ \emph {et~al.}(2011)\citenamefont
  {Kasztelan}, \citenamefont {Trotzky}, \citenamefont {Chen}, \citenamefont
  {Bloch}, \citenamefont {McCulloch}, \citenamefont {Schollw\"ock},\ and\
  \citenamefont {Orso}}]{PhysRevLett.106.155302}%
  \BibitemOpen
  \bibfield  {author} {\bibinfo {author} {\bibfnamefont {C.}~\bibnamefont
  {Kasztelan}}, \bibinfo {author} {\bibfnamefont {S.}~\bibnamefont {Trotzky}},
  \bibinfo {author} {\bibfnamefont {Y.-A.}\ \bibnamefont {Chen}}, \bibinfo
  {author} {\bibfnamefont {I.}~\bibnamefont {Bloch}}, \bibinfo {author}
  {\bibfnamefont {I.~P.}\ \bibnamefont {McCulloch}}, \bibinfo {author}
  {\bibfnamefont {U.}~\bibnamefont {Schollw\"ock}}, \ and\ \bibinfo {author}
  {\bibfnamefont {G.}~\bibnamefont {Orso}},\ }\href {\doibase
  10.1103/PhysRevLett.106.155302} {\bibfield  {journal} {\bibinfo  {journal}
  {Phys. Rev. Lett.}\ }\textbf {\bibinfo {volume} {106}},\ \bibinfo {pages}
  {155302} (\bibinfo {year} {2011})}\BibitemShut {NoStop}%
\bibitem [{\citenamefont {An}\ \emph {et~al.}(2018)\citenamefont {An},
  \citenamefont {Meier}, \citenamefont {Ang'ong'a},\ and\ \citenamefont
  {Gadway}}]{PhysRevLett.120.040407}%
  \BibitemOpen
  \bibfield  {author} {\bibinfo {author} {\bibfnamefont {F.~A.}\ \bibnamefont
  {An}}, \bibinfo {author} {\bibfnamefont {E.~J.}\ \bibnamefont {Meier}},
  \bibinfo {author} {\bibfnamefont {J.}~\bibnamefont {Ang'ong'a}}, \ and\
  \bibinfo {author} {\bibfnamefont {B.}~\bibnamefont {Gadway}},\ }\href
  {\doibase 10.1103/PhysRevLett.120.040407} {\bibfield  {journal} {\bibinfo
  {journal} {Phys. Rev. Lett.}\ }\textbf {\bibinfo {volume} {120}},\ \bibinfo
  {pages} {040407} (\bibinfo {year} {2018})}\BibitemShut {NoStop}%
\bibitem [{\citenamefont {Zhang}\ \emph {et~al.}(2019)\citenamefont {Zhang},
  \citenamefont {Gui},\ and\ \citenamefont {Chen}}]{PhysRevA.99.023616}%
  \BibitemOpen
  \bibfield  {author} {\bibinfo {author} {\bibfnamefont {Y.}~\bibnamefont
  {Zhang}}, \bibinfo {author} {\bibfnamefont {Z.}~\bibnamefont {Gui}}, \ and\
  \bibinfo {author} {\bibfnamefont {Y.}~\bibnamefont {Chen}},\ }\href {\doibase
  10.1103/PhysRevA.99.023616} {\bibfield  {journal} {\bibinfo  {journal} {Phys.
  Rev. A}\ }\textbf {\bibinfo {volume} {99}},\ \bibinfo {pages} {023616}
  (\bibinfo {year} {2019})}\BibitemShut {NoStop}%
\bibitem [{\citenamefont {Guan}\ \emph {et~al.}(2020)\citenamefont {Guan},
  \citenamefont {Ome}, \citenamefont {Bersano}, \citenamefont {Mossman},
  \citenamefont {Engels},\ and\ \citenamefont
  {Blume}}]{PhysRevLett.125.213401}%
  \BibitemOpen
  \bibfield  {author} {\bibinfo {author} {\bibfnamefont {Q.}~\bibnamefont
  {Guan}}, \bibinfo {author} {\bibfnamefont {M.~K.~H.}\ \bibnamefont {Ome}},
  \bibinfo {author} {\bibfnamefont {T.~M.}\ \bibnamefont {Bersano}}, \bibinfo
  {author} {\bibfnamefont {S.}~\bibnamefont {Mossman}}, \bibinfo {author}
  {\bibfnamefont {P.}~\bibnamefont {Engels}}, \ and\ \bibinfo {author}
  {\bibfnamefont {D.}~\bibnamefont {Blume}},\ }\href {\doibase
  10.1103/PhysRevLett.125.213401} {\bibfield  {journal} {\bibinfo  {journal}
  {Phys. Rev. Lett.}\ }\textbf {\bibinfo {volume} {125}},\ \bibinfo {pages}
  {213401} (\bibinfo {year} {2020})}\BibitemShut {NoStop}%
\bibitem [{\citenamefont {Sinitsyn}(2002)}]{PhysRevB.66.205303}%
  \BibitemOpen
  \bibfield  {author} {\bibinfo {author} {\bibfnamefont {N.~A.}\ \bibnamefont
  {Sinitsyn}},\ }\href {\doibase 10.1103/PhysRevB.66.205303} {\bibfield
  {journal} {\bibinfo  {journal} {Phys. Rev. B}\ }\textbf {\bibinfo {volume}
  {66}},\ \bibinfo {pages} {205303} (\bibinfo {year} {2002})}\BibitemShut
  {NoStop}%
\bibitem [{\citenamefont {Shytov}(2004)}]{PhysRevA.70.052708}%
  \BibitemOpen
  \bibfield  {author} {\bibinfo {author} {\bibfnamefont {A.~V.}\ \bibnamefont
  {Shytov}},\ }\href {\doibase 10.1103/PhysRevA.70.052708} {\bibfield
  {journal} {\bibinfo  {journal} {Phys. Rev. A}\ }\textbf {\bibinfo {volume}
  {70}},\ \bibinfo {pages} {052708} (\bibinfo {year} {2004})}\BibitemShut
  {NoStop}%
\bibitem [{\citenamefont {Volkov}\ and\ \citenamefont
  {Ostrovsky}(2004)}]{Volkov_2004}%
  \BibitemOpen
  \bibfield  {author} {\bibinfo {author} {\bibfnamefont {M.~V.}\ \bibnamefont
  {Volkov}}\ and\ \bibinfo {author} {\bibfnamefont {V.~N.}\ \bibnamefont
  {Ostrovsky}},\ }\href {\doibase 10.1088/0953-4075/37/20/003} {\bibfield
  {journal} {\bibinfo  {journal} {J. Phys. B: At. Mol. Opt. Phys.}\ }\textbf
  {\bibinfo {volume} {37}},\ \bibinfo {pages} {4069} (\bibinfo {year}
  {2004})}\BibitemShut {NoStop}%
\bibitem [{\citenamefont {Wang}\ \emph {et~al.}(2006)\citenamefont {Wang},
  \citenamefont {Ye}, \citenamefont {Fu}, \citenamefont {Chen},\ and\
  \citenamefont {Liu}}]{PhysRevA.74.033414}%
  \BibitemOpen
  \bibfield  {author} {\bibinfo {author} {\bibfnamefont {G.-F.}\ \bibnamefont
  {Wang}}, \bibinfo {author} {\bibfnamefont {D.-F.}\ \bibnamefont {Ye}},
  \bibinfo {author} {\bibfnamefont {L.-B.}\ \bibnamefont {Fu}}, \bibinfo
  {author} {\bibfnamefont {X.-Z.}\ \bibnamefont {Chen}}, \ and\ \bibinfo
  {author} {\bibfnamefont {J.}~\bibnamefont {Liu}},\ }\href {\doibase
  10.1103/PhysRevA.74.033414} {\bibfield  {journal} {\bibinfo  {journal} {Phys.
  Rev. A}\ }\textbf {\bibinfo {volume} {74}},\ \bibinfo {pages} {033414}
  (\bibinfo {year} {2006})}\BibitemShut {NoStop}%
\bibitem [{\citenamefont {Sinitsyn}(2013)}]{PhysRevA.87.032701}%
  \BibitemOpen
  \bibfield  {author} {\bibinfo {author} {\bibfnamefont {N.~A.}\ \bibnamefont
  {Sinitsyn}},\ }\href {\doibase 10.1103/PhysRevA.87.032701} {\bibfield
  {journal} {\bibinfo  {journal} {Phys. Rev. A}\ }\textbf {\bibinfo {volume}
  {87}},\ \bibinfo {pages} {032701} (\bibinfo {year} {2013})}\BibitemShut
  {NoStop}%
\bibitem [{\citenamefont {Sinitsyn}(2014)}]{PhysRevA.90.062509}%
  \BibitemOpen
  \bibfield  {author} {\bibinfo {author} {\bibfnamefont {N.~A.}\ \bibnamefont
  {Sinitsyn}},\ }\href {\doibase 10.1103/PhysRevA.90.062509} {\bibfield
  {journal} {\bibinfo  {journal} {Phys. Rev. A}\ }\textbf {\bibinfo {volume}
  {90}},\ \bibinfo {pages} {062509} (\bibinfo {year} {2014})}\BibitemShut
  {NoStop}%
\bibitem [{\citenamefont {Ashhab}(2016)}]{PhysRevA.94.042109}%
  \BibitemOpen
  \bibfield  {author} {\bibinfo {author} {\bibfnamefont {S.}~\bibnamefont
  {Ashhab}},\ }\href {\doibase 10.1103/PhysRevA.94.042109} {\bibfield
  {journal} {\bibinfo  {journal} {Phys. Rev. A}\ }\textbf {\bibinfo {volume}
  {94}},\ \bibinfo {pages} {042109} (\bibinfo {year} {2016})}\BibitemShut
  {NoStop}%
\bibitem [{\citenamefont {Li}\ \emph {et~al.}(2017)\citenamefont {Li},
  \citenamefont {Sun}, \citenamefont {Chernyak},\ and\ \citenamefont
  {Sinitsyn}}]{PhysRevA.96.022107}%
  \BibitemOpen
  \bibfield  {author} {\bibinfo {author} {\bibfnamefont {F.}~\bibnamefont
  {Li}}, \bibinfo {author} {\bibfnamefont {C.}~\bibnamefont {Sun}}, \bibinfo
  {author} {\bibfnamefont {V.~Y.}\ \bibnamefont {Chernyak}}, \ and\ \bibinfo
  {author} {\bibfnamefont {N.~A.}\ \bibnamefont {Sinitsyn}},\ }\href {\doibase
  10.1103/PhysRevA.96.022107} {\bibfield  {journal} {\bibinfo  {journal} {Phys.
  Rev. A}\ }\textbf {\bibinfo {volume} {96}},\ \bibinfo {pages} {022107}
  (\bibinfo {year} {2017})}\BibitemShut {NoStop}%
\bibitem [{\citenamefont {Malla}\ and\ \citenamefont
  {Raikh}(2017)}]{PhysRevB.96.115437}%
  \BibitemOpen
  \bibfield  {author} {\bibinfo {author} {\bibfnamefont {R.~K.}\ \bibnamefont
  {Malla}}\ and\ \bibinfo {author} {\bibfnamefont {M.~E.}\ \bibnamefont
  {Raikh}},\ }\href {\doibase 10.1103/PhysRevB.96.115437} {\bibfield  {journal}
  {\bibinfo  {journal} {Phys. Rev. B}\ }\textbf {\bibinfo {volume} {96}},\
  \bibinfo {pages} {115437} (\bibinfo {year} {2017})}\BibitemShut {NoStop}%
\bibitem [{\citenamefont {Garanin}\ and\ \citenamefont
  {Schilling}(2002)}]{PhysRevB.66.174438}%
  \BibitemOpen
  \bibfield  {author} {\bibinfo {author} {\bibfnamefont {D.~A.}\ \bibnamefont
  {Garanin}}\ and\ \bibinfo {author} {\bibfnamefont {R.}~\bibnamefont
  {Schilling}},\ }\href {\doibase 10.1103/PhysRevB.66.174438} {\bibfield
  {journal} {\bibinfo  {journal} {Phys. Rev. B}\ }\textbf {\bibinfo {volume}
  {66}},\ \bibinfo {pages} {174438} (\bibinfo {year} {2002})}\BibitemShut
  {NoStop}%
\bibitem [{\citenamefont {Dou}\ \emph {et~al.}(2018)\citenamefont {Dou},
  \citenamefont {Liu},\ and\ \citenamefont {Fu}}]{PhysRevA.98.022102}%
  \BibitemOpen
  \bibfield  {author} {\bibinfo {author} {\bibfnamefont {F.-Q.}\ \bibnamefont
  {Dou}}, \bibinfo {author} {\bibfnamefont {J.}~\bibnamefont {Liu}}, \ and\
  \bibinfo {author} {\bibfnamefont {L.-B.}\ \bibnamefont {Fu}},\ }\href
  {\doibase 10.1103/PhysRevA.98.022102} {\bibfield  {journal} {\bibinfo
  {journal} {Phys. Rev. A}\ }\textbf {\bibinfo {volume} {98}},\ \bibinfo
  {pages} {022102} (\bibinfo {year} {2018})}\BibitemShut {NoStop}%
\bibitem [{\citenamefont {Bender}\ and\ \citenamefont
  {Boettcher}(1998)}]{PhysRevLett.80.5243}%
  \BibitemOpen
  \bibfield  {author} {\bibinfo {author} {\bibfnamefont {C.~M.}\ \bibnamefont
  {Bender}}\ and\ \bibinfo {author} {\bibfnamefont {S.}~\bibnamefont
  {Boettcher}},\ }\href {\doibase 10.1103/PhysRevLett.80.5243} {\bibfield
  {journal} {\bibinfo  {journal} {Phys. Rev. Lett.}\ }\textbf {\bibinfo
  {volume} {80}},\ \bibinfo {pages} {5243} (\bibinfo {year}
  {1998})}\BibitemShut {NoStop}%
\bibitem [{\citenamefont {Konotop}\ \emph {et~al.}(2016)\citenamefont
  {Konotop}, \citenamefont {Yang},\ and\ \citenamefont
  {Zezyulin}}]{RevModPhys.88.035002}%
  \BibitemOpen
  \bibfield  {author} {\bibinfo {author} {\bibfnamefont {V.~V.}\ \bibnamefont
  {Konotop}}, \bibinfo {author} {\bibfnamefont {J.}~\bibnamefont {Yang}}, \
  and\ \bibinfo {author} {\bibfnamefont {D.~A.}\ \bibnamefont {Zezyulin}},\
  }\href {\doibase 10.1103/RevModPhys.88.035002} {\bibfield  {journal}
  {\bibinfo  {journal} {Rev. Mod. Phys.}\ }\textbf {\bibinfo {volume} {88}},\
  \bibinfo {pages} {035002} (\bibinfo {year} {2016})}\BibitemShut {NoStop}%
\bibitem [{\citenamefont {Feng}\ \emph {et~al.}(2017)\citenamefont {Feng},
  \citenamefont {El-Ganainy},\ and\ \citenamefont {Ge}}]{Feng2017}%
  \BibitemOpen
  \bibfield  {author} {\bibinfo {author} {\bibfnamefont {L.}~\bibnamefont
  {Feng}}, \bibinfo {author} {\bibfnamefont {R.}~\bibnamefont {El-Ganainy}}, \
  and\ \bibinfo {author} {\bibfnamefont {L.}~\bibnamefont {Ge}},\ }\href
  {\doibase 10.1038/s41566-017-0031-1} {\bibfield  {journal} {\bibinfo
  {journal} {Nat. Photon.}\ }\textbf {\bibinfo {volume} {11}},\ \bibinfo
  {pages} {752} (\bibinfo {year} {2017})}\BibitemShut {NoStop}%
\bibitem [{\citenamefont {Miri}\ and\ \citenamefont
  {Al{\`u}}(2019)}]{Mirieaar7709}%
  \BibitemOpen
  \bibfield  {author} {\bibinfo {author} {\bibfnamefont {M.-A.}\ \bibnamefont
  {Miri}}\ and\ \bibinfo {author} {\bibfnamefont {A.}~\bibnamefont {Al{\`u}}},\
  }\href {\doibase 10.1126/science.aar7709} {\bibfield  {journal} {\bibinfo
  {journal} {Science}\ }\textbf {\bibinfo {volume} {363}},\ \bibinfo {pages}
  {6422} (\bibinfo {year} {2019})}\BibitemShut {NoStop}%
\bibitem [{\citenamefont {El-Ganainy}\ \emph {et~al.}(2018)\citenamefont
  {El-Ganainy}, \citenamefont {Makris}, \citenamefont {Khajavikhan},
  \citenamefont {Musslimani}, \citenamefont {Rotter},\ and\ \citenamefont
  {Christodoulides}}]{El-Ganainy2018}%
  \BibitemOpen
  \bibfield  {author} {\bibinfo {author} {\bibfnamefont {R.}~\bibnamefont
  {El-Ganainy}}, \bibinfo {author} {\bibfnamefont {K.~G.}\ \bibnamefont
  {Makris}}, \bibinfo {author} {\bibfnamefont {M.}~\bibnamefont {Khajavikhan}},
  \bibinfo {author} {\bibfnamefont {Z.~H.}\ \bibnamefont {Musslimani}},
  \bibinfo {author} {\bibfnamefont {S.}~\bibnamefont {Rotter}}, \ and\ \bibinfo
  {author} {\bibfnamefont {D.~N.}\ \bibnamefont {Christodoulides}},\ }\href
  {\doibase 10.1038/nphys4323} {\bibfield  {journal} {\bibinfo  {journal} {Nat.
  Phys.}\ }\textbf {\bibinfo {volume} {14}},\ \bibinfo {pages} {11} (\bibinfo
  {year} {2018})}\BibitemShut {NoStop}%
\bibitem [{\citenamefont {Miri}\ and\ \citenamefont
  {Al\`{u}}(2019)}]{doi:10.1126/science.aar7709}%
  \BibitemOpen
  \bibfield  {author} {\bibinfo {author} {\bibfnamefont {M.-A.}\ \bibnamefont
  {Miri}}\ and\ \bibinfo {author} {\bibfnamefont {A.}~\bibnamefont {Al\`{u}}},\
  }\href {\doibase 10.1126/science.aar7709} {\bibfield  {journal} {\bibinfo
  {journal} {Science}\ }\textbf {\bibinfo {volume} {363}},\ \bibinfo {pages}
  {eaar7709} (\bibinfo {year} {2019})}\BibitemShut {NoStop}%
\bibitem [{\citenamefont {Bouganne}\ \emph {et~al.}(2020)\citenamefont
  {Bouganne}, \citenamefont {Bosch~Aguilera}, \citenamefont {Ghermaoui},
  \citenamefont {Beugnon},\ and\ \citenamefont {Gerbier}}]{Bouganne2020}%
  \BibitemOpen
  \bibfield  {author} {\bibinfo {author} {\bibfnamefont {R.}~\bibnamefont
  {Bouganne}}, \bibinfo {author} {\bibfnamefont {M.}~\bibnamefont
  {Bosch~Aguilera}}, \bibinfo {author} {\bibfnamefont {A.}~\bibnamefont
  {Ghermaoui}}, \bibinfo {author} {\bibfnamefont {J.}~\bibnamefont {Beugnon}},
  \ and\ \bibinfo {author} {\bibfnamefont {F.}~\bibnamefont {Gerbier}},\ }\href
  {\doibase 10.1038/s41567-019-0678-2} {\bibfield  {journal} {\bibinfo
  {journal} {Nat. Phys.}\ }\textbf {\bibinfo {volume} {16}},\ \bibinfo {pages}
  {21} (\bibinfo {year} {2020})}\BibitemShut {NoStop}%
\bibitem [{\citenamefont {Ashida}\ \emph {et~al.}(2020)\citenamefont {Ashida},
  \citenamefont {Gong},\ and\ \citenamefont
  {Ueda}}]{doi:10.1080/00018732.2021.1876991}%
  \BibitemOpen
  \bibfield  {author} {\bibinfo {author} {\bibfnamefont {Y.}~\bibnamefont
  {Ashida}}, \bibinfo {author} {\bibfnamefont {Z.}~\bibnamefont {Gong}}, \ and\
  \bibinfo {author} {\bibfnamefont {M.}~\bibnamefont {Ueda}},\ }\href {\doibase
  10.1080/00018732.2021.1876991} {\bibfield  {journal} {\bibinfo  {journal}
  {Adv. Phys.}\ }\textbf {\bibinfo {volume} {69}},\ \bibinfo {pages} {249}
  (\bibinfo {year} {2020})}\BibitemShut {NoStop}%
\bibitem [{\citenamefont {Bergholtz}\ \emph {et~al.}(2021)\citenamefont
  {Bergholtz}, \citenamefont {Budich},\ and\ \citenamefont
  {Kunst}}]{RevModPhys.93.015005}%
  \BibitemOpen
  \bibfield  {author} {\bibinfo {author} {\bibfnamefont {E.~J.}\ \bibnamefont
  {Bergholtz}}, \bibinfo {author} {\bibfnamefont {J.~C.}\ \bibnamefont
  {Budich}}, \ and\ \bibinfo {author} {\bibfnamefont {F.~K.}\ \bibnamefont
  {Kunst}},\ }\href {\doibase 10.1103/RevModPhys.93.015005} {\bibfield
  {journal} {\bibinfo  {journal} {Rev. Mod. Phys.}\ }\textbf {\bibinfo {volume}
  {93}},\ \bibinfo {pages} {015005} (\bibinfo {year} {2021})}\BibitemShut
  {NoStop}%
\bibitem [{\citenamefont {Gefen}\ \emph {et~al.}(1987)\citenamefont {Gefen},
  \citenamefont {Ben-Jacob},\ and\ \citenamefont
  {Caldeira}}]{PhysRevB.36.2770}%
  \BibitemOpen
  \bibfield  {author} {\bibinfo {author} {\bibfnamefont {Y.}~\bibnamefont
  {Gefen}}, \bibinfo {author} {\bibfnamefont {E.}~\bibnamefont {Ben-Jacob}}, \
  and\ \bibinfo {author} {\bibfnamefont {A.~O.}\ \bibnamefont {Caldeira}},\
  }\href {\doibase 10.1103/PhysRevB.36.2770} {\bibfield  {journal} {\bibinfo
  {journal} {Phys. Rev. B}\ }\textbf {\bibinfo {volume} {36}},\ \bibinfo
  {pages} {2770} (\bibinfo {year} {1987})}\BibitemShut {NoStop}%
\bibitem [{\citenamefont {Akulin}\ and\ \citenamefont
  {Schleich}(1992)}]{PhysRevA.46.4110}%
  \BibitemOpen
  \bibfield  {author} {\bibinfo {author} {\bibfnamefont {V.~M.}\ \bibnamefont
  {Akulin}}\ and\ \bibinfo {author} {\bibfnamefont {W.~P.}\ \bibnamefont
  {Schleich}},\ }\href {\doibase 10.1103/PhysRevA.46.4110} {\bibfield
  {journal} {\bibinfo  {journal} {Phys. Rev. A}\ }\textbf {\bibinfo {volume}
  {46}},\ \bibinfo {pages} {4110} (\bibinfo {year} {1992})}\BibitemShut
  {NoStop}%
\bibitem [{\citenamefont {Avishai}\ and\ \citenamefont
  {Band}(2014)}]{PhysRevA.90.032116}%
  \BibitemOpen
  \bibfield  {author} {\bibinfo {author} {\bibfnamefont {Y.}~\bibnamefont
  {Avishai}}\ and\ \bibinfo {author} {\bibfnamefont {Y.~B.}\ \bibnamefont
  {Band}},\ }\href {\doibase 10.1103/PhysRevA.90.032116} {\bibfield  {journal}
  {\bibinfo  {journal} {Phys. Rev. A}\ }\textbf {\bibinfo {volume} {90}},\
  \bibinfo {pages} {032116} (\bibinfo {year} {2014})}\BibitemShut {NoStop}%
\bibitem [{\citenamefont {He}\ and\ \citenamefont
  {Jones}(2021)}]{PhysRevA.104.013111}%
  \BibitemOpen
  \bibfield  {author} {\bibinfo {author} {\bibfnamefont {C.}~\bibnamefont
  {He}}\ and\ \bibinfo {author} {\bibfnamefont {R.~R.}\ \bibnamefont {Jones}},\
  }\href {\doibase 10.1103/PhysRevA.104.013111} {\bibfield  {journal} {\bibinfo
   {journal} {Phys. Rev. A}\ }\textbf {\bibinfo {volume} {104}},\ \bibinfo
  {pages} {013111} (\bibinfo {year} {2021})}\BibitemShut {NoStop}%
\bibitem [{\citenamefont {Sounas}\ and\ \citenamefont
  {Al\`{u}}(2017)}]{Sounas2017}%
  \BibitemOpen
  \bibfield  {author} {\bibinfo {author} {\bibfnamefont {D.~L.}\ \bibnamefont
  {Sounas}}\ and\ \bibinfo {author} {\bibfnamefont {A.}~\bibnamefont
  {Al\`{u}}},\ }\href {\doibase 10.1038/s41566-017-0051-x} {\bibfield
  {journal} {\bibinfo  {journal} {Nat. Photon.}\ }\textbf {\bibinfo {volume}
  {12}},\ \bibinfo {pages} {774} (\bibinfo {year} {2017})}\BibitemShut
  {NoStop}%
\bibitem [{\citenamefont {Fruchart}\ \emph {et~al.}(2021)\citenamefont
  {Fruchart}, \citenamefont {Hanai}, \citenamefont {Littlewood},\ and\
  \citenamefont {Vitelli}}]{Fruchart2021}%
  \BibitemOpen
  \bibfield  {author} {\bibinfo {author} {\bibfnamefont {M.}~\bibnamefont
  {Fruchart}}, \bibinfo {author} {\bibfnamefont {R.}~\bibnamefont {Hanai}},
  \bibinfo {author} {\bibfnamefont {P.~B.}\ \bibnamefont {Littlewood}}, \ and\
  \bibinfo {author} {\bibfnamefont {V.}~\bibnamefont {Vitelli}},\ }\href
  {\doibase 10.1038/s41586-021-03375-9} {\bibfield  {journal} {\bibinfo
  {journal} {Nature}\ }\textbf {\bibinfo {volume} {592}},\ \bibinfo {pages}
  {363} (\bibinfo {year} {2021})}\BibitemShut {NoStop}%
\bibitem [{\citenamefont {Castin}\ and\ \citenamefont
  {Dum}(1997)}]{PhysRevLett.79.3553}%
  \BibitemOpen
  \bibfield  {author} {\bibinfo {author} {\bibfnamefont {Y.}~\bibnamefont
  {Castin}}\ and\ \bibinfo {author} {\bibfnamefont {R.}~\bibnamefont {Dum}},\
  }\href {\doibase 10.1103/PhysRevLett.79.3553} {\bibfield  {journal} {\bibinfo
   {journal} {Phys. Rev. Lett.}\ }\textbf {\bibinfo {volume} {79}},\ \bibinfo
  {pages} {3553} (\bibinfo {year} {1997})}\BibitemShut {NoStop}%
\bibitem [{\citenamefont {Castin}\ and\ \citenamefont
  {Dum}(1998)}]{PhysRevA.57.3008}%
  \BibitemOpen
  \bibfield  {author} {\bibinfo {author} {\bibfnamefont {Y.}~\bibnamefont
  {Castin}}\ and\ \bibinfo {author} {\bibfnamefont {R.}~\bibnamefont {Dum}},\
  }\href {\doibase 10.1103/PhysRevA.57.3008} {\bibfield  {journal} {\bibinfo
  {journal} {Phys. Rev. A}\ }\textbf {\bibinfo {volume} {57}},\ \bibinfo
  {pages} {3008} (\bibinfo {year} {1998})}\BibitemShut {NoStop}%
\bibitem [{\citenamefont {Liu}\ \emph {et~al.}(2006)\citenamefont {Liu},
  \citenamefont {Zhang}, \citenamefont {Raizen},\ and\ \citenamefont
  {Niu}}]{PhysRevA.73.013601}%
  \BibitemOpen
  \bibfield  {author} {\bibinfo {author} {\bibfnamefont {J.}~\bibnamefont
  {Liu}}, \bibinfo {author} {\bibfnamefont {C.}~\bibnamefont {Zhang}}, \bibinfo
  {author} {\bibfnamefont {M.~G.}\ \bibnamefont {Raizen}}, \ and\ \bibinfo
  {author} {\bibfnamefont {Q.}~\bibnamefont {Niu}},\ }\href {\doibase
  10.1103/PhysRevA.73.013601} {\bibfield  {journal} {\bibinfo  {journal} {Phys.
  Rev. A}\ }\textbf {\bibinfo {volume} {73}},\ \bibinfo {pages} {013601}
  (\bibinfo {year} {2006})}\BibitemShut {NoStop}%
\bibitem [{\citenamefont {Su}\ \emph {et~al.}(1980)\citenamefont {Su},
  \citenamefont {Schrieffer},\ and\ \citenamefont {Heeger}}]{PhysRevB.22.2099}%
  \BibitemOpen
  \bibfield  {author} {\bibinfo {author} {\bibfnamefont {W.~P.}\ \bibnamefont
  {Su}}, \bibinfo {author} {\bibfnamefont {J.~R.}\ \bibnamefont {Schrieffer}},
  \ and\ \bibinfo {author} {\bibfnamefont {A.~J.}\ \bibnamefont {Heeger}},\
  }\href {\doibase 10.1103/PhysRevB.22.2099} {\bibfield  {journal} {\bibinfo
  {journal} {Phys. Rev. B}\ }\textbf {\bibinfo {volume} {22}},\ \bibinfo
  {pages} {2099} (\bibinfo {year} {1980})}\BibitemShut {NoStop}%
\bibitem [{\citenamefont {Wang}\ \emph {et~al.}(2009)\citenamefont {Wang},
  \citenamefont {Chong}, \citenamefont {Joannopoulos},\ and\ \citenamefont
  {Solja\u{c}i\'{c}}}]{Wang2009}%
  \BibitemOpen
  \bibfield  {author} {\bibinfo {author} {\bibfnamefont {Z.}~\bibnamefont
  {Wang}}, \bibinfo {author} {\bibfnamefont {Y.}~\bibnamefont {Chong}},
  \bibinfo {author} {\bibfnamefont {J.~D.}\ \bibnamefont {Joannopoulos}}, \
  and\ \bibinfo {author} {\bibfnamefont {M.}~\bibnamefont {Solja\u{c}i\'{c}}},\
  }\href {\doibase 10.1038/nature08293} {\bibfield  {journal} {\bibinfo
  {journal} {Nature}\ }\textbf {\bibinfo {volume} {461}},\ \bibinfo {pages}
  {772} (\bibinfo {year} {2009})}\BibitemShut {NoStop}%
\bibitem [{\citenamefont {Qi}\ and\ \citenamefont
  {Zhang}(2010)}]{doi:10.1063/1.3293411}%
  \BibitemOpen
  \bibfield  {author} {\bibinfo {author} {\bibfnamefont {X.-L.}\ \bibnamefont
  {Qi}}\ and\ \bibinfo {author} {\bibfnamefont {S.-C.}\ \bibnamefont {Zhang}},\
  }\href {\doibase 10.1063/1.3293411} {\bibfield  {journal} {\bibinfo
  {journal} {Physics Today}\ }\textbf {\bibinfo {volume} {63}},\ \bibinfo
  {pages} {33} (\bibinfo {year} {2010})}\BibitemShut {NoStop}%
\bibitem [{\citenamefont {Lodahl}\ \emph {et~al.}(2017)\citenamefont {Lodahl},
  \citenamefont {Mahmoodian}, \citenamefont {Stobbe}, \citenamefont
  {Rauschenbeutel}, \citenamefont {Schneeweiss}, \citenamefont {Volz},
  \citenamefont {Pichler},\ and\ \citenamefont {Zoller}}]{Lodahl2017}%
  \BibitemOpen
  \bibfield  {author} {\bibinfo {author} {\bibfnamefont {P.}~\bibnamefont
  {Lodahl}}, \bibinfo {author} {\bibfnamefont {S.}~\bibnamefont {Mahmoodian}},
  \bibinfo {author} {\bibfnamefont {S.}~\bibnamefont {Stobbe}}, \bibinfo
  {author} {\bibfnamefont {A.}~\bibnamefont {Rauschenbeutel}}, \bibinfo
  {author} {\bibfnamefont {P.}~\bibnamefont {Schneeweiss}}, \bibinfo {author}
  {\bibfnamefont {J.}~\bibnamefont {Volz}}, \bibinfo {author} {\bibfnamefont
  {H.}~\bibnamefont {Pichler}}, \ and\ \bibinfo {author} {\bibfnamefont
  {P.}~\bibnamefont {Zoller}},\ }\href {\doibase 10.1038/nature21037}
  {\bibfield  {journal} {\bibinfo  {journal} {Nature}\ }\textbf {\bibinfo
  {volume} {541}},\ \bibinfo {pages} {473} (\bibinfo {year}
  {2017})}\BibitemShut {NoStop}%
\bibitem [{\citenamefont {Stannigel}\ \emph {et~al.}(2012)\citenamefont
  {Stannigel}, \citenamefont {Komar}, \citenamefont {Habraken}, \citenamefont
  {Bennett}, \citenamefont {Lukin}, \citenamefont {Zoller},\ and\ \citenamefont
  {Rabl}}]{PhysRevLett.109.013603}%
  \BibitemOpen
  \bibfield  {author} {\bibinfo {author} {\bibfnamefont {K.}~\bibnamefont
  {Stannigel}}, \bibinfo {author} {\bibfnamefont {P.}~\bibnamefont {Komar}},
  \bibinfo {author} {\bibfnamefont {S.~J.~M.}\ \bibnamefont {Habraken}},
  \bibinfo {author} {\bibfnamefont {S.~D.}\ \bibnamefont {Bennett}}, \bibinfo
  {author} {\bibfnamefont {M.~D.}\ \bibnamefont {Lukin}}, \bibinfo {author}
  {\bibfnamefont {P.}~\bibnamefont {Zoller}}, \ and\ \bibinfo {author}
  {\bibfnamefont {P.}~\bibnamefont {Rabl}},\ }\href {\doibase
  10.1103/PhysRevLett.109.013603} {\bibfield  {journal} {\bibinfo  {journal}
  {Phys. Rev. Lett.}\ }\textbf {\bibinfo {volume} {109}},\ \bibinfo {pages}
  {013603} (\bibinfo {year} {2012})}\BibitemShut {NoStop}%
\bibitem [{\citenamefont {Kim}\ \emph {et~al.}(2015)\citenamefont {Kim},
  \citenamefont {Kuzyk}, \citenamefont {Han}, \citenamefont {Wang},\ and\
  \citenamefont {Bahl}}]{Kim2015}%
  \BibitemOpen
  \bibfield  {author} {\bibinfo {author} {\bibfnamefont {J.}~\bibnamefont
  {Kim}}, \bibinfo {author} {\bibfnamefont {M.~C.}\ \bibnamefont {Kuzyk}},
  \bibinfo {author} {\bibfnamefont {K.}~\bibnamefont {Han}}, \bibinfo {author}
  {\bibfnamefont {H.}~\bibnamefont {Wang}}, \ and\ \bibinfo {author}
  {\bibfnamefont {G.}~\bibnamefont {Bahl}},\ }\href {\doibase
  10.1038/nphys3236} {\bibfield  {journal} {\bibinfo  {journal} {Nat. Phys.}\
  }\textbf {\bibinfo {volume} {11}},\ \bibinfo {pages} {275} (\bibinfo {year}
  {2015})}\BibitemShut {NoStop}%
\bibitem [{\citenamefont {Barzanjeh}\ \emph {et~al.}(2018)\citenamefont
  {Barzanjeh}, \citenamefont {Aquilina},\ and\ \citenamefont
  {Xuereb}}]{PhysRevLett.120.060601}%
  \BibitemOpen
  \bibfield  {author} {\bibinfo {author} {\bibfnamefont {S.}~\bibnamefont
  {Barzanjeh}}, \bibinfo {author} {\bibfnamefont {M.}~\bibnamefont {Aquilina}},
  \ and\ \bibinfo {author} {\bibfnamefont {A.}~\bibnamefont {Xuereb}},\ }\href
  {\doibase 10.1103/PhysRevLett.120.060601} {\bibfield  {journal} {\bibinfo
  {journal} {Phys. Rev. Lett.}\ }\textbf {\bibinfo {volume} {120}},\ \bibinfo
  {pages} {060601} (\bibinfo {year} {2018})}\BibitemShut {NoStop}%
\bibitem [{\citenamefont {Li}\ \emph {et~al.}(2018)\citenamefont {Li},
  \citenamefont {Kottos},\ and\ \citenamefont
  {Shapiro}}]{PhysRevApplied.9.044031}%
  \BibitemOpen
  \bibfield  {author} {\bibinfo {author} {\bibfnamefont {H.}~\bibnamefont
  {Li}}, \bibinfo {author} {\bibfnamefont {T.}~\bibnamefont {Kottos}}, \ and\
  \bibinfo {author} {\bibfnamefont {B.}~\bibnamefont {Shapiro}},\ }\href
  {\doibase 10.1103/PhysRevApplied.9.044031} {\bibfield  {journal} {\bibinfo
  {journal} {Phys. Rev. Applied}\ }\textbf {\bibinfo {volume} {9}},\ \bibinfo
  {pages} {044031} (\bibinfo {year} {2018})}\BibitemShut {NoStop}%
\bibitem [{\citenamefont {Gong}\ \emph {et~al.}(2018)\citenamefont {Gong},
  \citenamefont {Ashida}, \citenamefont {Kawabata}, \citenamefont {Takasan},
  \citenamefont {Higashikawa},\ and\ \citenamefont {Ueda}}]{PhysRevX.8.031079}%
  \BibitemOpen
  \bibfield  {author} {\bibinfo {author} {\bibfnamefont {Z.}~\bibnamefont
  {Gong}}, \bibinfo {author} {\bibfnamefont {Y.}~\bibnamefont {Ashida}},
  \bibinfo {author} {\bibfnamefont {K.}~\bibnamefont {Kawabata}}, \bibinfo
  {author} {\bibfnamefont {K.}~\bibnamefont {Takasan}}, \bibinfo {author}
  {\bibfnamefont {S.}~\bibnamefont {Higashikawa}}, \ and\ \bibinfo {author}
  {\bibfnamefont {M.}~\bibnamefont {Ueda}},\ }\href {\doibase
  10.1103/PhysRevX.8.031079} {\bibfield  {journal} {\bibinfo  {journal} {Phys.
  Rev. X}\ }\textbf {\bibinfo {volume} {8}},\ \bibinfo {pages} {031079}
  (\bibinfo {year} {2018})}\BibitemShut {NoStop}%
\bibitem [{\citenamefont {Zhang}\ \emph {et~al.}(2018)\citenamefont {Zhang},
  \citenamefont {Zhu}, \citenamefont {Zhao}, \citenamefont {Yan},\ and\
  \citenamefont {Zhu}}]{doi:10.1080/00018732.2019.1594094}%
  \BibitemOpen
  \bibfield  {author} {\bibinfo {author} {\bibfnamefont {D.-W.}\ \bibnamefont
  {Zhang}}, \bibinfo {author} {\bibfnamefont {Y.-Q.}\ \bibnamefont {Zhu}},
  \bibinfo {author} {\bibfnamefont {Y.~X.}\ \bibnamefont {Zhao}}, \bibinfo
  {author} {\bibfnamefont {H.}~\bibnamefont {Yan}}, \ and\ \bibinfo {author}
  {\bibfnamefont {S.-L.}\ \bibnamefont {Zhu}},\ }\href {\doibase
  10.1080/00018732.2019.1594094} {\bibfield  {journal} {\bibinfo  {journal}
  {Adv. Phys.}\ }\textbf {\bibinfo {volume} {67}},\ \bibinfo {pages} {253}
  (\bibinfo {year} {2018})}\BibitemShut {NoStop}%
\bibitem [{\citenamefont {Gou}\ \emph {et~al.}(2020)\citenamefont {Gou},
  \citenamefont {Chen}, \citenamefont {Xie}, \citenamefont {Xiao},
  \citenamefont {Deng}, \citenamefont {Gadway}, \citenamefont {Yi},\ and\
  \citenamefont {Yan}}]{PhysRevLett.124.070402}%
  \BibitemOpen
  \bibfield  {author} {\bibinfo {author} {\bibfnamefont {W.}~\bibnamefont
  {Gou}}, \bibinfo {author} {\bibfnamefont {T.}~\bibnamefont {Chen}}, \bibinfo
  {author} {\bibfnamefont {D.}~\bibnamefont {Xie}}, \bibinfo {author}
  {\bibfnamefont {T.}~\bibnamefont {Xiao}}, \bibinfo {author} {\bibfnamefont
  {T.-S.}\ \bibnamefont {Deng}}, \bibinfo {author} {\bibfnamefont
  {B.}~\bibnamefont {Gadway}}, \bibinfo {author} {\bibfnamefont
  {W.}~\bibnamefont {Yi}}, \ and\ \bibinfo {author} {\bibfnamefont
  {B.}~\bibnamefont {Yan}},\ }\href {\doibase 10.1103/PhysRevLett.124.070402}
  {\bibfield  {journal} {\bibinfo  {journal} {Phys. Rev. Lett.}\ }\textbf
  {\bibinfo {volume} {124}},\ \bibinfo {pages} {070402} (\bibinfo {year}
  {2020})}\BibitemShut {NoStop}%
\bibitem [{\citenamefont {Born}(1927)}]{Born1927}%
  \BibitemOpen
  \bibfield  {author} {\bibinfo {author} {\bibfnamefont {M.}~\bibnamefont
  {Born}},\ }\href {\doibase 10.1007/BF01400360} {\bibfield  {journal}
  {\bibinfo  {journal} {Z. Phys.}\ }\textbf {\bibinfo {volume} {40}},\ \bibinfo
  {pages} {167} (\bibinfo {year} {1927})}\BibitemShut {NoStop}%
\bibitem [{\citenamefont {Born}\ and\ \citenamefont {Fock}(1928)}]{Born1928}%
  \BibitemOpen
  \bibfield  {author} {\bibinfo {author} {\bibfnamefont {M.}~\bibnamefont
  {Born}}\ and\ \bibinfo {author} {\bibfnamefont {V.}~\bibnamefont {Fock}},\
  }\href {\doibase 10.1007/BF01343193} {\bibfield  {journal} {\bibinfo
  {journal} {Z. Phys.}\ }\textbf {\bibinfo {volume} {51}},\ \bibinfo {pages}
  {165} (\bibinfo {year} {1928})}\BibitemShut {NoStop}%
\bibitem [{\citenamefont {Farhi}\ \emph {et~al.}(2000)\citenamefont {Farhi},
  \citenamefont {Goldstone}, \citenamefont {Gutmann},\ and\ \citenamefont
  {Sipser}}]{quant-ph/0001106}%
  \BibitemOpen
  \bibfield  {author} {\bibinfo {author} {\bibfnamefont {E.}~\bibnamefont
  {Farhi}}, \bibinfo {author} {\bibfnamefont {J.}~\bibnamefont {Goldstone}},
  \bibinfo {author} {\bibfnamefont {S.}~\bibnamefont {Gutmann}}, \ and\
  \bibinfo {author} {\bibfnamefont {M.}~\bibnamefont {Sipser}},\ }\href@noop {}
  {\enquote {\bibinfo {title} {Quantum computation by adiabatic evolution},}\ }
  (\bibinfo {year} {2000}),\ \Eprint
  {http://arxiv.org/abs/arXiv:quant-ph/0001106} {arXiv:quant-ph/0001106}
  \BibitemShut {NoStop}%
\bibitem [{\citenamefont {Farhi}\ \emph {et~al.}(2001)\citenamefont {Farhi},
  \citenamefont {Goldstone}, \citenamefont {Gutmann}, \citenamefont {Lapan},
  \citenamefont {Lundgren},\ and\ \citenamefont
  {Preda}}]{doi:10.1126/science.1057726}%
  \BibitemOpen
  \bibfield  {author} {\bibinfo {author} {\bibfnamefont {E.}~\bibnamefont
  {Farhi}}, \bibinfo {author} {\bibfnamefont {J.}~\bibnamefont {Goldstone}},
  \bibinfo {author} {\bibfnamefont {S.}~\bibnamefont {Gutmann}}, \bibinfo
  {author} {\bibfnamefont {J.}~\bibnamefont {Lapan}}, \bibinfo {author}
  {\bibfnamefont {A.}~\bibnamefont {Lundgren}}, \ and\ \bibinfo {author}
  {\bibfnamefont {D.}~\bibnamefont {Preda}},\ }\href {\doibase
  10.1126/science.1057726} {\bibfield  {journal} {\bibinfo  {journal}
  {Science}\ }\textbf {\bibinfo {volume} {292}},\ \bibinfo {pages} {472}
  (\bibinfo {year} {2001})}\BibitemShut {NoStop}%
\bibitem [{\citenamefont {Albash}\ and\ \citenamefont
  {Lidar}(2018)}]{RevModPhys.90.015002}%
  \BibitemOpen
  \bibfield  {author} {\bibinfo {author} {\bibfnamefont {T.}~\bibnamefont
  {Albash}}\ and\ \bibinfo {author} {\bibfnamefont {D.~A.}\ \bibnamefont
  {Lidar}},\ }\href {\doibase 10.1103/RevModPhys.90.015002} {\bibfield
  {journal} {\bibinfo  {journal} {Rev. Mod. Phys.}\ }\textbf {\bibinfo {volume}
  {90}},\ \bibinfo {pages} {015002} (\bibinfo {year} {2018})}\BibitemShut
  {NoStop}%
\bibitem [{\citenamefont {Gu\'ery-Odelin}\ \emph {et~al.}(2019)\citenamefont
  {Gu\'ery-Odelin}, \citenamefont {Ruschhaupt}, \citenamefont {Kiely},
  \citenamefont {Torrontegui}, \citenamefont {Mart\'{\i}nez-Garaot},\ and\
  \citenamefont {Muga}}]{RevModPhys.91.045001}%
  \BibitemOpen
  \bibfield  {author} {\bibinfo {author} {\bibfnamefont {D.}~\bibnamefont
  {Gu\'ery-Odelin}}, \bibinfo {author} {\bibfnamefont {A.}~\bibnamefont
  {Ruschhaupt}}, \bibinfo {author} {\bibfnamefont {A.}~\bibnamefont {Kiely}},
  \bibinfo {author} {\bibfnamefont {E.}~\bibnamefont {Torrontegui}}, \bibinfo
  {author} {\bibfnamefont {S.}~\bibnamefont {Mart\'{\i}nez-Garaot}}, \ and\
  \bibinfo {author} {\bibfnamefont {J.~G.}\ \bibnamefont {Muga}},\ }\href
  {\doibase 10.1103/RevModPhys.91.045001} {\bibfield  {journal} {\bibinfo
  {journal} {Rev. Mod. Phys.}\ }\textbf {\bibinfo {volume} {91}},\ \bibinfo
  {pages} {045001} (\bibinfo {year} {2019})}\BibitemShut {NoStop}%
\bibitem [{\citenamefont {Liu}\ and\ \citenamefont
  {Fu}(2010)}]{PhysRevA.81.052112}%
  \BibitemOpen
  \bibfield  {author} {\bibinfo {author} {\bibfnamefont {J.}~\bibnamefont
  {Liu}}\ and\ \bibinfo {author} {\bibfnamefont {L.~B.}\ \bibnamefont {Fu}},\
  }\href {\doibase 10.1103/PhysRevA.81.052112} {\bibfield  {journal} {\bibinfo
  {journal} {Phys. Rev. A}\ }\textbf {\bibinfo {volume} {81}},\ \bibinfo
  {pages} {052112} (\bibinfo {year} {2010})}\BibitemShut {NoStop}%
\bibitem [{\citenamefont {Wu}\ \emph {et~al.}(2005)\citenamefont {Wu},
  \citenamefont {Liu},\ and\ \citenamefont {Niu}}]{PhysRevLett.94.140402}%
  \BibitemOpen
  \bibfield  {author} {\bibinfo {author} {\bibfnamefont {B.}~\bibnamefont
  {Wu}}, \bibinfo {author} {\bibfnamefont {J.}~\bibnamefont {Liu}}, \ and\
  \bibinfo {author} {\bibfnamefont {Q.}~\bibnamefont {Niu}},\ }\href {\doibase
  10.1103/PhysRevLett.94.140402} {\bibfield  {journal} {\bibinfo  {journal}
  {Phys. Rev. Lett.}\ }\textbf {\bibinfo {volume} {94}},\ \bibinfo {pages}
  {140402} (\bibinfo {year} {2005})}\BibitemShut {NoStop}%
\bibitem [{\citenamefont {Pu}\ \emph {et~al.}(2007)\citenamefont {Pu},
  \citenamefont {Maenner}, \citenamefont {Zhang},\ and\ \citenamefont
  {Ling}}]{PhysRevLett.98.050406}%
  \BibitemOpen
  \bibfield  {author} {\bibinfo {author} {\bibfnamefont {H.}~\bibnamefont
  {Pu}}, \bibinfo {author} {\bibfnamefont {P.}~\bibnamefont {Maenner}},
  \bibinfo {author} {\bibfnamefont {W.}~\bibnamefont {Zhang}}, \ and\ \bibinfo
  {author} {\bibfnamefont {H.~Y.}\ \bibnamefont {Ling}},\ }\href {\doibase
  10.1103/PhysRevLett.98.050406} {\bibfield  {journal} {\bibinfo  {journal}
  {Phys. Rev. Lett.}\ }\textbf {\bibinfo {volume} {98}},\ \bibinfo {pages}
  {050406} (\bibinfo {year} {2007})}\BibitemShut {NoStop}%
\bibitem [{\citenamefont {Liu}\ \emph {et~al.}(2018)\citenamefont {Liu},
  \citenamefont {Li}, \citenamefont {Fu},\ and\ \citenamefont {Ye}}]{Liu2018}%
  \BibitemOpen
  \bibfield  {author} {\bibinfo {author} {\bibfnamefont {J.}~\bibnamefont
  {Liu}}, \bibinfo {author} {\bibfnamefont {S.~C.}\ \bibnamefont {Li}},
  \bibinfo {author} {\bibfnamefont {L.~B.}\ \bibnamefont {Fu}}, \ and\ \bibinfo
  {author} {\bibfnamefont {D.~F.}\ \bibnamefont {Ye}},\ }\href {\doibase
  10.1007/978-981-13-2643-1_2} {\emph {\bibinfo {title} {Nonlinear Adiabatic
  Evolution of Quantum Systems}}}\ (\bibinfo  {publisher} {Springer,
  Singapore},\ \bibinfo {year} {2018})\BibitemShut {NoStop}%
\bibitem [{\citenamefont {Ib\'a\~nez}\ \emph {et~al.}(2011)\citenamefont
  {Ib\'a\~nez}, \citenamefont {Mart\'{\i}nez-Garaot}, \citenamefont {Chen},
  \citenamefont {Torrontegui},\ and\ \citenamefont
  {Muga}}]{PhysRevA.84.023415}%
  \BibitemOpen
  \bibfield  {author} {\bibinfo {author} {\bibfnamefont {S.}~\bibnamefont
  {Ib\'a\~nez}}, \bibinfo {author} {\bibfnamefont {S.}~\bibnamefont
  {Mart\'{\i}nez-Garaot}}, \bibinfo {author} {\bibfnamefont {X.}~\bibnamefont
  {Chen}}, \bibinfo {author} {\bibfnamefont {E.}~\bibnamefont {Torrontegui}}, \
  and\ \bibinfo {author} {\bibfnamefont {J.~G.}\ \bibnamefont {Muga}},\ }\href
  {\doibase 10.1103/PhysRevA.84.023415} {\bibfield  {journal} {\bibinfo
  {journal} {Phys. Rev. A}\ }\textbf {\bibinfo {volume} {84}},\ \bibinfo
  {pages} {023415} (\bibinfo {year} {2011})}\BibitemShut {NoStop}%
\bibitem [{\citenamefont {Zhang}\ and\ \citenamefont
  {Wu}(2019)}]{PhysRevA.99.032121}%
  \BibitemOpen
  \bibfield  {author} {\bibinfo {author} {\bibfnamefont {Q.}~\bibnamefont
  {Zhang}}\ and\ \bibinfo {author} {\bibfnamefont {B.}~\bibnamefont {Wu}},\
  }\href {\doibase 10.1103/PhysRevA.99.032121} {\bibfield  {journal} {\bibinfo
  {journal} {Phys. Rev. A}\ }\textbf {\bibinfo {volume} {99}},\ \bibinfo
  {pages} {032121} (\bibinfo {year} {2019})}\BibitemShut {NoStop}%
\bibitem [{\citenamefont {Luan}\ \emph {et~al.}(2022)\citenamefont {Luan},
  \citenamefont {Shen},\ and\ \citenamefont {Yi}}]{PhysRevA.105.013714}%
  \BibitemOpen
  \bibfield  {author} {\bibinfo {author} {\bibfnamefont {T.~Z.}\ \bibnamefont
  {Luan}}, \bibinfo {author} {\bibfnamefont {H.~Z.}\ \bibnamefont {Shen}}, \
  and\ \bibinfo {author} {\bibfnamefont {X.~X.}\ \bibnamefont {Yi}},\ }\href
  {\doibase 10.1103/PhysRevA.105.013714} {\bibfield  {journal} {\bibinfo
  {journal} {Phys. Rev. A}\ }\textbf {\bibinfo {volume} {105}},\ \bibinfo
  {pages} {013714} (\bibinfo {year} {2022})}\BibitemShut {NoStop}%
\bibitem [{\citenamefont {Eckel}\ \emph {et~al.}(2014)\citenamefont {Eckel},
  \citenamefont {Lee}, \citenamefont {Jendrzejewski}, \citenamefont {Murray},
  \citenamefont {Clark}, \citenamefont {Lobb}, \citenamefont {Phillips},
  \citenamefont {Edwards},\ and\ \citenamefont {Campbell}}]{Eckel2014}%
  \BibitemOpen
  \bibfield  {author} {\bibinfo {author} {\bibfnamefont {S.}~\bibnamefont
  {Eckel}}, \bibinfo {author} {\bibfnamefont {J.~G.}\ \bibnamefont {Lee}},
  \bibinfo {author} {\bibfnamefont {F.}~\bibnamefont {Jendrzejewski}}, \bibinfo
  {author} {\bibfnamefont {N.}~\bibnamefont {Murray}}, \bibinfo {author}
  {\bibfnamefont {C.~W.}\ \bibnamefont {Clark}}, \bibinfo {author}
  {\bibfnamefont {C.~J.}\ \bibnamefont {Lobb}}, \bibinfo {author}
  {\bibfnamefont {W.~D.}\ \bibnamefont {Phillips}}, \bibinfo {author}
  {\bibfnamefont {M.}~\bibnamefont {Edwards}}, \ and\ \bibinfo {author}
  {\bibfnamefont {G.~K.}\ \bibnamefont {Campbell}},\ }\href {\doibase
  10.1038/nature12958} {\bibfield  {journal} {\bibinfo  {journal} {Nature}\
  }\textbf {\bibinfo {volume} {506}},\ \bibinfo {pages} {200} (\bibinfo {year}
  {2014})}\BibitemShut {NoStop}%
\bibitem [{\citenamefont {Anderson}(1964)}]{Anderson1963}%
  \BibitemOpen
  \bibfield  {author} {\bibinfo {author} {\bibfnamefont {P.~W.}\ \bibnamefont
  {Anderson}},\ }in\ \href@noop {} {\emph {\bibinfo {booktitle} {Fifth
  International Spring School of Physics}}},\ \bibinfo {editor} {edited by\
  \bibinfo {editor} {\bibfnamefont {E.~R.}\ \bibnamefont {Caianello}}}\
  (\bibinfo  {publisher} {Academic},\ \bibinfo {address} {New York},\ \bibinfo
  {year} {1964})\BibitemShut {NoStop}%
\bibitem [{\citenamefont {Shimshoni}\ \emph {et~al.}(1989)\citenamefont
  {Shimshoni}, \citenamefont {Gefen},\ and\ \citenamefont
  {Fishman}}]{PhysRevB.40.2158}%
  \BibitemOpen
  \bibfield  {author} {\bibinfo {author} {\bibfnamefont {E.}~\bibnamefont
  {Shimshoni}}, \bibinfo {author} {\bibfnamefont {Y.}~\bibnamefont {Gefen}}, \
  and\ \bibinfo {author} {\bibfnamefont {S.}~\bibnamefont {Fishman}},\ }\href
  {\doibase 10.1103/PhysRevB.40.2158} {\bibfield  {journal} {\bibinfo
  {journal} {Phys. Rev. B}\ }\textbf {\bibinfo {volume} {40}},\ \bibinfo
  {pages} {2158} (\bibinfo {year} {1989})}\BibitemShut {NoStop}%
\bibitem [{\citenamefont {Smerzi}\ \emph {et~al.}(1997)\citenamefont {Smerzi},
  \citenamefont {Fantoni}, \citenamefont {Giovanazzi},\ and\ \citenamefont
  {Shenoy}}]{PhysRevLett.79.4950}%
  \BibitemOpen
  \bibfield  {author} {\bibinfo {author} {\bibfnamefont {A.}~\bibnamefont
  {Smerzi}}, \bibinfo {author} {\bibfnamefont {S.}~\bibnamefont {Fantoni}},
  \bibinfo {author} {\bibfnamefont {S.}~\bibnamefont {Giovanazzi}}, \ and\
  \bibinfo {author} {\bibfnamefont {S.~R.}\ \bibnamefont {Shenoy}},\ }\href
  {\doibase 10.1103/PhysRevLett.79.4950} {\bibfield  {journal} {\bibinfo
  {journal} {Phys. Rev. Lett.}\ }\textbf {\bibinfo {volume} {79}},\ \bibinfo
  {pages} {4950} (\bibinfo {year} {1997})}\BibitemShut {NoStop}%
\bibitem [{\citenamefont {Berry}(1994)}]{Berry1994}%
  \BibitemOpen
  \bibfield  {author} {\bibinfo {author} {\bibfnamefont {M.}~\bibnamefont
  {Berry}},\ }\href {\doibase http://www.jstor.org/stable/24095727} {\bibfield
  {journal} {\bibinfo  {journal} {Current Science}\ }\textbf {\bibinfo {volume}
  {67}},\ \bibinfo {pages} {220} (\bibinfo {year} {1994})}\BibitemShut
  {NoStop}%
\bibitem [{\citenamefont {Zhao}\ \emph {et~al.}(2020)\citenamefont {Zhao},
  \citenamefont {Zhou}, \citenamefont {Liu}, \citenamefont {Tong},\ and\
  \citenamefont {Huang}}]{PhysRevA.102.062213}%
  \BibitemOpen
  \bibfield  {author} {\bibinfo {author} {\bibfnamefont {W.-L.}\ \bibnamefont
  {Zhao}}, \bibinfo {author} {\bibfnamefont {L.}~\bibnamefont {Zhou}}, \bibinfo
  {author} {\bibfnamefont {J.}~\bibnamefont {Liu}}, \bibinfo {author}
  {\bibfnamefont {P.}~\bibnamefont {Tong}}, \ and\ \bibinfo {author}
  {\bibfnamefont {K.}~\bibnamefont {Huang}},\ }\href {\doibase
  10.1103/PhysRevA.102.062213} {\bibfield  {journal} {\bibinfo  {journal}
  {Phys. Rev. A}\ }\textbf {\bibinfo {volume} {102}},\ \bibinfo {pages}
  {062213} (\bibinfo {year} {2020})}\BibitemShut {NoStop}%
\bibitem [{\citenamefont {Zhou}\ \emph {et~al.}(2018)\citenamefont {Zhou},
  \citenamefont {Wang}, \citenamefont {Wang},\ and\ \citenamefont
  {Gong}}]{PhysRevA.98.022129}%
  \BibitemOpen
  \bibfield  {author} {\bibinfo {author} {\bibfnamefont {L.}~\bibnamefont
  {Zhou}}, \bibinfo {author} {\bibfnamefont {Q.-h.}\ \bibnamefont {Wang}},
  \bibinfo {author} {\bibfnamefont {H.}~\bibnamefont {Wang}}, \ and\ \bibinfo
  {author} {\bibfnamefont {J.}~\bibnamefont {Gong}},\ }\href {\doibase
  10.1103/PhysRevA.98.022129} {\bibfield  {journal} {\bibinfo  {journal} {Phys.
  Rev. A}\ }\textbf {\bibinfo {volume} {98}},\ \bibinfo {pages} {022129}
  (\bibinfo {year} {2018})}\BibitemShut {NoStop}%
\bibitem [{\citenamefont {Zhou}\ and\ \citenamefont
  {Gong}(2018)}]{PhysRevB.98.205417}%
  \BibitemOpen
  \bibfield  {author} {\bibinfo {author} {\bibfnamefont {L.}~\bibnamefont
  {Zhou}}\ and\ \bibinfo {author} {\bibfnamefont {J.}~\bibnamefont {Gong}},\
  }\href {\doibase 10.1103/PhysRevB.98.205417} {\bibfield  {journal} {\bibinfo
  {journal} {Phys. Rev. B}\ }\textbf {\bibinfo {volume} {98}},\ \bibinfo
  {pages} {205417} (\bibinfo {year} {2018})}\BibitemShut {NoStop}%
\bibitem [{\citenamefont {Longhi}(2021)}]{PhysRevB.103.054203}%
  \BibitemOpen
  \bibfield  {author} {\bibinfo {author} {\bibfnamefont {S.}~\bibnamefont
  {Longhi}},\ }\href {\doibase 10.1103/PhysRevB.103.054203} {\bibfield
  {journal} {\bibinfo  {journal} {Phys. Rev. B}\ }\textbf {\bibinfo {volume}
  {103}},\ \bibinfo {pages} {054203} (\bibinfo {year} {2021})}\BibitemShut
  {NoStop}%
\end{thebibliography}
%\begin{thebibliography}{100}
%\end{thebibliography}
%

\end{document}